\documentclass{aa}   

\usepackage{rotating, threeparttable}
\usepackage{graphicx}
\usepackage{tabularx}
\usepackage{booktabs}
\usepackage{rotating}
\usepackage{float}
\usepackage{orcidlink}
\usepackage{txfonts}
\usepackage{placeins}
\begin{document}

   \title{Periodic activity on the (sub)mm surface of the AGB star Mira A}

   \author{M.~Andriantsaralaza\inst{1}\orcidlink{0000-0001-5539-4316}
          \and
        B.~Bojnordi~Arbab\inst{1}\orcidlink{0000-0002-6901-4086}
         \and
        W.~Vlemmings\inst{1}\orcidlink{0000-0002-2700-9916}
         \and
        T.~Khouri\inst{1}\orcidlink{0000-0002-5680-9525}
         \and 
        L.~Planquart\inst{1}\orcidlink{0000-0003-0392-6645}
         \and 
        K.~T.~Wong\inst{2}\orcidlink{0000-0002-4579-6546}
        \and
        J.~Alcolea\inst{3}
        \and
        E.~Humphreys\inst{4}
         \and
        S.~del~Palacio\inst{1}\orcidlink{0000-0002-5761-2417}
        \and 
        M.~Wittkowski\inst{4}   
        }

   \institute{Department of Physics and Astronomy, Chalmers University of Technology, 412 96, Gothenburg, Sweden\\\email{miora.andriantsaralaza@chalmers.se}
    \and
    Department of Physics and Astronomy, Uppsala University, Box 524, 751 20 Uppsala, Sweden 
    \and
    Observatorio Astronómico Nacional (OAN-IGN), Alfonso XII 3, 28014 Madrid, Spain
    \and
    European Southern Observatory (ESO), Karl-Schwarzschild-Straße 2, 85748, Garching bei München, Germany
}

   \date{Received 2026/06/09; accepted 2026/10/13}

  \abstract
    {Evolved stars on the asymptotic giant branch (AGB) lose mass through stellar winds launched from their dynamic extended atmospheres. Recent high-angular-resolution observations of the archetypal AGB star Mira~A ($o$~Ceti) have revealed compact hotspots on its (sub)millimetre surface, whose origin remains unclear.}
    {We aim to characterise the (sub)millimetre photosphere of Mira~A and investigate the properties and temporal behaviour of the compact hotspots detected on its surface.}
    {We analysed 19 epochs of high-angular-resolution observations obtained with the Atacama Large Millimetre/submillimetre Array in Bands~4, 6, 7, and~9, covering frequencies between $\sim$\,131.30 and 669.25\,GHz. The complex visibilities were fitted with parametric models to derive the flux density, size, morphology, and brightness temperature of the stellar disc and an additional more compact component across multiple pulsation phases and observing frequencies.}
    {Our results show that, near maximum light, the flux density of the stellar disc increases with frequency while its apparent size decreases, in agreement with the expected stratified radio photosphere. In Band~6, both the flux density and the size of the stellar disc appear to vary with pulsation phase, with the largest measured values occurring close to optical maximum light and the lowest observed around minimum light.
    Compact hotspots are detected in Bands~6 and 7 near maximum light, with their brightness temperatures exceeding those of the stellar disc by several thousand kelvin and reaching a temperature excess of $\sim 8200$\,K in Band~6. They are highly energetic and are detected over timescales of weeks, while the available observations also suggest potential longer-term changes in both their position and intensity.}
    {Hotspot activity appears to occur repeatedly across multiple pulsation cycles and may be linked to strong shocks, magnetic activity, and/or episodic mass ejections, and could therefore play an important role in the mass-loss process in AGB stars.}

   \keywords{Stars: AGB and post-AGB -- Stars: activity --
                Stars: atmospheres --
                Stars: individual: $o$~Ceti
               }

   \maketitle

\section{Introduction}
Asymptotic Giant Branch (AGB) stars are major contributors to the chemical enrichment of the interstellar medium thanks to their characteristic massive stellar winds. These winds are thought to be driven by radiation pressure on dust grains, which form at a few stellar radii from the star \citep{HoefnerOlof2018}. The material that forms this dust originates in the dynamic extended atmosphere, where pulsation, convection, and shocks lift the gas outward. Studying this region is, therefore, essential for understanding the mass-loss process in AGB stars.

Recent high-angular-resolution interferometric observations have revealed asymmetries, variability, and localised surface activity in the extended atmospheres of nearby AGB stars \citep[e.g. R~Dor, W~Hya, R~Leo, CW~Leo;][]{Matthews2015, Vlemmings2015, Wong2016, Matthews2018, Ohnaka2025}. Imaging of stars such as $\pi^1$~Gruis and R~Dor has uncovered regions of enhanced and reduced intensity relative to the mean surface brightness, consistent with convective motions \citep[e.g.][]{Paladini2018, Ohnaka2019}, with a characteristic timescale of about one month \citep{Vlemmings2024}. In addition, a few AGB stars, including W~Hya, R~Dor, and Mira~A, exhibit compact hotspots at millimetre/submillimetre (mm/submm) wavelengths that are significantly hotter than the surrounding stellar surface \citep{Vlemmings2015, Vlemmings2017, Vlemmings2019}. These hotspots are more compact and energetic than the convective cells predicted by current 3D atmosphere models \citep[e.g.][]{Freytag2017, Freytag2023}. Their origin and role in AGB atmospheres remain unclear, motivating further investigation.

Mira is a binary system with a white dwarf companion at a current projected separation of ~0.4 arcsec \citep{Khouri2026}. Mira~A ($o$~Ceti), the archetypal Mira-type variable with a pulsation period of 331.3 days \citep{Watson2006}, is one of the nearest ($\sim$\,100\,pc) and well-studied AGB stars \citep[e.g.][]{Haniff1995, vanLeeuwen2007, Wong2016, Planesas2016, Perrin2020}. Its proximity and large angular diameter make it an ideal target for resolving the structure and variability of its extended atmosphere. In this work, we analysed 19 epochs of observations from the Atacama Large Millimetre/submillimetre Array (ALMA) to investigate the temporal and spectral variability of the stellar disc and compact hotspots of Mira~A. The multi-epoch and multi-frequency coverage in this work provides an opportunity to gain insights into their evolution and to probe emission arising from different atmospheric depths.
This paper is organised as follows. In Sect.~\ref{sec: obs and analysis}, we present the observations used in this analysis and describe the data reduction process. Section~\ref{sec: visibility fitting} outlines the visibility-fitting method used to derive the stellar properties. Finally, in Sect.~\ref{sec: results}, we present and discuss the resulting constraints on the stellar properties and activity across the (sub)mm surface of Mira~A.

\section{Observations and data reduction}
\label{sec: obs and analysis}

We retrieved ALMA observations of the Mira system with nominal angular resolutions $\leq 60$ mas from the ALMA archive, spanning 10 ALMA projects, with some of them having multiple observing epochs. We define an epoch as observations obtained on the same calendar day. Multiple executions of the same scheduling block on a given day are therefore treated as a single epoch, whereas observations with the same setup obtained on different days are considered separate epochs. Observations previously analysed for the same purpose as this work, i.e. those in which the stellar disc and hotspots of the AGB star Mira A were characterised using identical visibility-fitting methods \citep[e.g.][]{Vlemmings2015}, were excluded. A detailed inspection of the data revealed several observations and/or spectral windows (spws) that had either bandwidths that were too narrow, as they were designed to target maser emission (spws 0\,--\,5 of ALMA project 2018.1.00551.S and spws 0\,--\,1 of ALMA project 2024.1.01776.S), too low velocity resolution ($\sim38$\,km\,s$^{-1}$) corresponding to observations performed in Time Division Mode (ALMA project 2016.1.01425.S), or jumps in their spectra (spw 3 of ALMA project 2016.1.00004.S) that prevented a sufficiently robust recovery of the continuum signal. These observations and/or spws were discarded. In addition, spws 1 and 2 of the Band~6 2017-10-06 epoch were also excluded from the subsequent analysis, as they were notably noisier and missing flux, with a recovered total flux value that is more than 30\% lower than the flux detected in spws 0 and 3. The final dataset consists of 19 epochs in Bands~4, 6, 7, and 9, with nearly all observations containing multiple spws of 1.875 GHz bandwidth each.

The raw measurement sets were calibrated using the standard Common Astronomy Software Applications (\textsc{CASA}) pipeline \citep{McMullin2007, CASA2022}. After an initial round of line flagging and imaging of the spectral cubes and channel-averaged continuum (to 40 or 20 channels), we performed a more systematic line-identification procedure using spectra extracted from cubes within apertures of 10, 20, and 40 pixels.
To identify spectral lines, we used the median absolute deviation (MAD) as a robust estimator of the data dispersion. Data points exceeding five times the MAD from the median were first flagged to remove strong outliers. The remaining distribution was then fitted with a Gaussian profile centred on the continuum level, and all points lying more than three standard deviations from this fit were subsequently flagged. This two-step line flagging procedure proved particularly important for the higher-frequency Bands 7 and 9, where strong maser features and weak, blended lines that artificially elevated the apparent continuum level hindered reliable continuum extraction. 

\begin{figure}[h]
\includegraphics[page=1,width=0.5\textwidth]{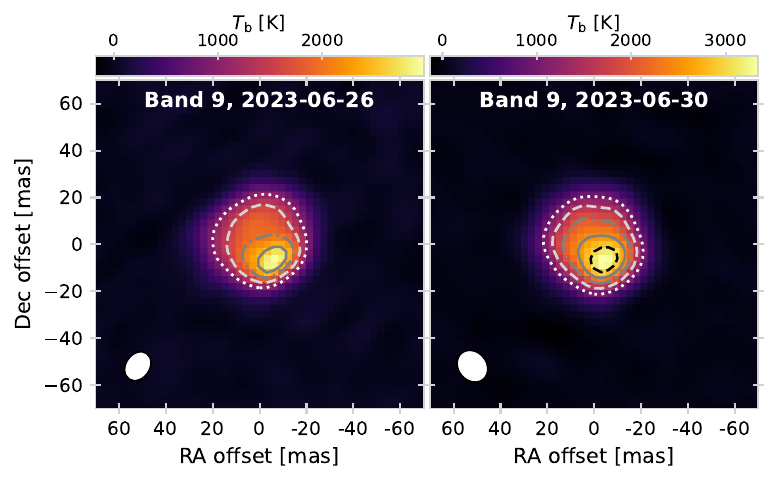}
\caption{ALMA Band 9 continuum map of Mira A in units of brightness temperature, $T_\mathrm{b}$ (see Eq.~\ref{eq: temp}), with a mean frequency of $\sim$\,666.24\,GHz, generated using superuniform weighting. The contours represent emission at 1000\,K (white dotted), 1500\,K (light gray dashed), 2000\,K (dark gray dot-dashed), 2500\,K (dark gray solid), and 3000\,K (black dashed). The white ellipse at the bottom-left corner of each image represents the synthesised beam (see Table~\ref{tab:observations}). The rms noise level is $\sim27.9$~K for the 2023-06-26 epoch and $\sim21.6$~K for the 2023-06-30 epoch.}
\label{fig: Band 9}
\end{figure}

We re-imaged the continuum data using Briggs weighting with a robust parameter of 0.5. We then performed one or two iterations of phase-only self-calibration on the stellar continuum, depending on the achieved improvement in the signal-to-noise ratio ($\mathrm{S/N}$), resulting in better dynamic range of the images by up to a factor of $\sim$\,$16$.  
Since the observations include both components of the Mira system, and our analysis focuses exclusively on Mira~A, the contribution from its companion, Mira~B, was removed prior to final imaging and analysis.  We ran the \textsc{CASA} task \textsc{tclean} with a circular mask centred on the position of Mira~B. The \textsc{clean} components were then subtracted from the visibilities using the task \textsc{uvsub}, thereby isolating Mira~A. The visibilities were subsequently recentred on Mira~A with the \textsc{CASA} task \textsc{fixvis}. High-angular-resolution continuum images of Mira~A were produced using superuniform visibility weighting, where the visibility-weight density is evaluated over a larger region of the uv-grid than in standard uniform weighting. This results in improved angular resolution compared to Briggs robust weighting schemes (e.g. a robust value of 0.5), but overall yields a lower rms noise level than a pure uniform weighting. In general, the dynamic ranges of the superuniform images were lower by a factor of $\leq$\,3 compared to images produced with Briggs weighting with a robust value of 0.5. The observing epochs, central frequencies, number of spws, corresponding ALMA project numbers, rms noise levels for the final aggregate continuum band, the sizes, and PAs of the synthesised beams from the superuniform and uniform (see Sect.~\ref{sec: visibility fitting} for their relevance in the analysis) images are summarised in Table~\ref{tab:observations}. The superuniform continuum images of Mira~A are shown in Figs.~\ref{fig: Band 9} (Band 9), \ref{fig: Continuum 2}, and \ref{fig: Continuum}.

\section{Visibility fitting}
\label{sec: visibility fitting}

We performed the analysis directly on the visibilities, making use of the over-resolution capability of interferometers such as ALMA in Fourier space, where even sources smaller than the nominal diffraction limit can be constrained, provided that the observations have sufficiently high sensitivity \citep[e.g.][]{MartiV2012}. The minimum measurable angular size from visibility fitting is defined as the smallest source size for which the departure of the visibilities from those of a point source remains statistically distinguishable from the noise level. It is given by
\begin{equation}
    s_\mathrm{lim} = \beta \, \Big ( \frac{\lambda_\mathrm{c}}{2 \, \mathrm{\mathrm{(S/N_\star)}}^2} \Big ) ^{1/4} \, \mathrm{FWHM}\,\, \mathrm{mas} \mathrm{,}
\label{Eq: size limit}
\end{equation}
where $\beta$ is a parameter that depends on the spatial distribution of the elements of the array, typically between 0.5 and 1; $\lambda_\mathrm{c}$ characterises the confidence of the size detection, as it depends on the probability of measuring a false size for a point-like source. In this work, we adopted conservative values of $\beta = 1$ and $\lambda_\mathrm{c} = 3.84$, corresponding to a 5\% chance of false-size detection. The FWHM corresponds to the full width at half maximum of the synthesised beam obtained using uniform weighting listed in Table~\ref{tab:observations}. The $\mathrm{S/N}$ of the weighted visibility average, $\mathrm{S/N}_\star$, is defined as $\langle V\rangle/\sigma$. For a compact, nearly unresolved source, the weighted visibility average can be approximated by the source flux density, $\langle V\rangle \simeq F_\nu$, while $\sigma$ is represented by the rms noise measured in the uniform-weighted image (see Table~\ref{tab:observations}). 

We fitted the complex visibilities using the \textsc{uvmultifit} code \citep{MartiV2014}. Each spw of each epoch was fitted separately. The parameter posterior distributions of each fitted parameter were obtained using a nested sampling approach (Bojnordi Arbab et al., in prep.) implemented with the \textsc{Dynesty} Python package \citep{Dynesty2020}. We adopted the median of the posterior distribution as the best-fit value for each parameter for each individual spw, with uncertainties defined by the 16th and 84th percentiles. For each epoch, the final parameter values were obtained by taking the median of the best-fit values across spws. The uncertainties were given by the larger of either the spw-to-spw spread relative to this median or the individual fitting uncertainties.
We followed this fitting procedure for two different scenarios: a uniform elliptical disc, and a disc plus a Gaussian component. 
Model selection was based primarily on the Bayesian evidence \citep{Kass1995}, but was also checked against the residual diagnostics in both the image and visibility domains. When the Bayesian evidence for a more complex model was not decisive and the residual diagnostics did not show a clear improvement over the simpler alternative, we conservatively adopted the simpler model.

We calculated brightness temperatures using
\begin{equation}
    T_\mathrm{b} = f  \frac{F_\nu}{\nu^2 \, \theta_\mathrm{min} \,\theta_\mathrm{maj}} \, \mathrm{K}
    \label{eq: temp}
\end{equation}
where $f = 1.76 \times 10^9$ for a disc, and $f = 1.22 \times 10^9$ for a Gaussian; $F_\nu$ is the flux density in mJy, $\theta_\mathrm{min}$ and $\theta_\mathrm{maj}$ are the minor and major axes of the modelled component in mas, respectively, and $v$ the central observing frequency in GHz. For the absolute flux density scale, we assumed a fractional error of 10\% for Bands 4, 6, and 7, and of 15\% for Band 9. The uncertainties in the derived brightness temperatures were obtained using a Monte Carlo approach, propagating the absolute flux-density uncertainties and the uncertainties in the fitted component sizes.

\section{Results and discussion}
\label{sec: results}
The best-fit model for most observations is a disc plus a Gaussian component, except for a few Band 6 epochs for which a disc model was clearly preferred (2017-10-06) or no robust detection of an additional component could be claimed (epochs 2017-10-04, 2017-10-05, and 2025-08-10). When a disc plus a Gaussian component was strongly favoured by the Bayesian evidence and produced a clear improvement in the residuals relative to a disc-only model, but the Gaussian component itself remained poorly constrained, we further tested a disc plus Dirac delta-function model. This revealed an intrinsic degeneracy in the inferred structure, particularly in Band~9, where different model configurations can reproduce the data comparably well (see Appendix~\ref{Appendix: subsection - degeneracy}). In such a case, we adopted a conservative approach and report the largest fitted sizes, such that the derived temperatures represent strict lower limits. 

For the Bands~6 observations obtained on 2017-09-22 (at a pulsation phase, $\phi$, of $\sim$\,0.67) and 2021-09-20 ($\phi \simeq 0.08$), the best-fit Gaussian component was found to be only marginally resolved (less than five times the minimum measurable scale, $s_\mathrm{lim}$) in one axis.  On the other hand, the fitted Gaussian components for the Band 7 2023-07-03 observation ($\phi \simeq 0.08$) and the Band 4 2017-09-21 ($\phi = 0.67$) epoch are unresolved ($< 3\,s_\mathrm{lim}$) in one or both directions. 
In such cases, when an axis is unresolved, we assumed a characteristic size corresponding to $3 s_\mathrm{lim}$ for that axis. In Table~\ref{tab:results} are listed the best-fit parameters of the best model: the flux density, the major axis diameter, the axis ratio (minor$/$major axis), and the position angle (PA) of the stellar disc and the Gaussian, when present, along with the $\mathrm{S/N}_\star$, the corresponding size limits, and the derived brightness temperatures (excess) for each epoch.

\subsection{The stellar disc}
\label{sec: discussion - radio photosphere}

\begin{figure*}[h]
\centering
\includegraphics[page=1,width=1\textwidth]{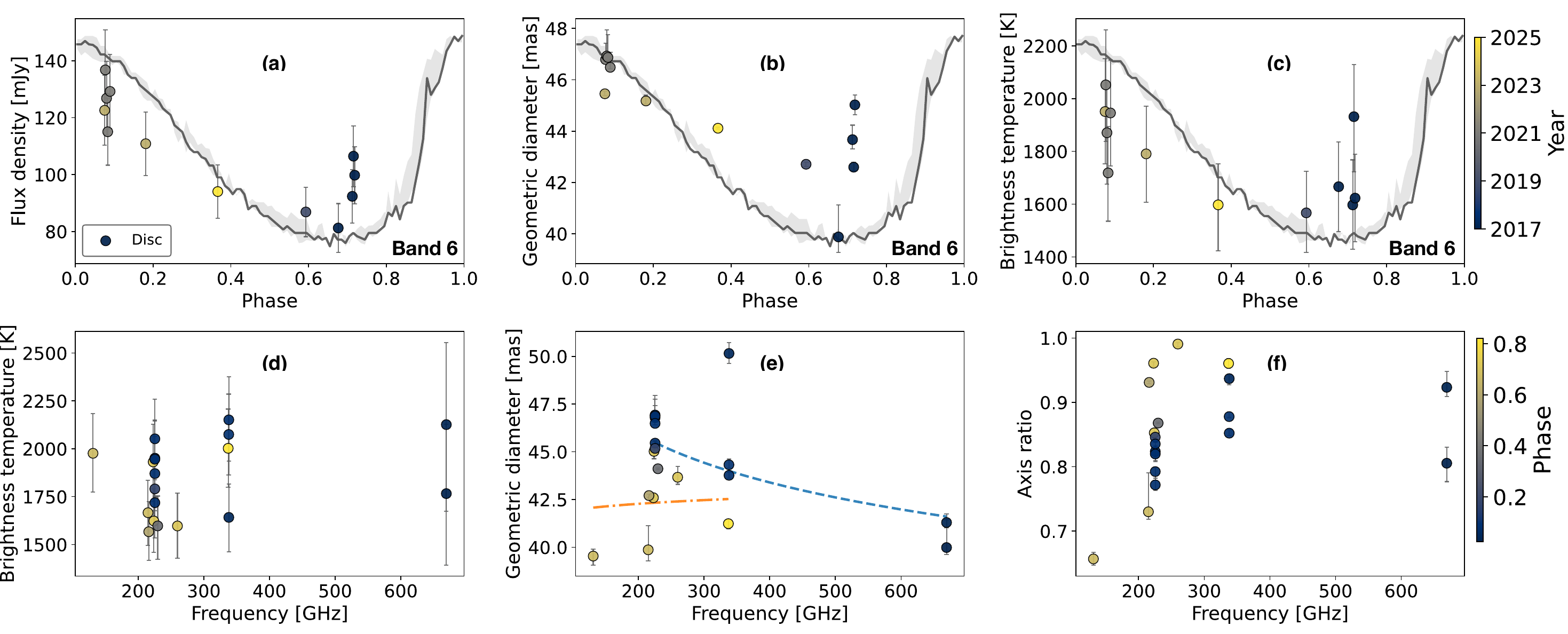}
\caption{Parameters of the fitted disc component: \textbf{(a)} Flux density of the disc as a function of pulsation phase in Band 6, scaled at a common frequency of 231~GHz using a spectral index of $\alpha=2$; \textbf{(b)} Geometric diameter of the disc in Band 6 as a function of phase; \textbf{(c)} Brightness temperature of the disc in Band 6 as a function of phase; \textbf{(d)} Disc brightness temperature versus observing frequency; \textbf{(e)} Geometric diameter of the disc against frequency, where the fit and slope of the size relation are given in blue (dashed line) for epochs within $0.0<\phi<0.2$ during the 2023 pulsation cycle and in orange (dashed-dotted line) for $0.6<\phi<0.9$ in the 2017 cycle (see text for details);  \textbf{(f)} Minor-to-Major axis ratio of the disc versus frequency. In all panels, one round symbol represents one observing epoch. The colours represent the year of observation in the phase plots (upper panels) and the pulsation phase in the frequency plots (lower panels). The solid line in all phase plots (upper panels) represents the median visual/\textit{V}-band light curve of Mira~A retrieved from the AAVSO (American Association of Variable Star Observers) database (\url{https://www.aavso.org/}) from November 2015 to September 2025, phase-folded using a reference epoch of $\mathrm{JD}_0 = 2458457.0$. The light curve was binned in phase with a bin width of 0.01, and the shaded region shows the 16th-84th percentile range of the visual magnitudes within each bin. The light curve was scaled with the presented parameters but originally spanned visual magnitudes from $\sim$~11 to 2.5~mag, with magnitude decreasing upward along the vertical axis.
}
\label{fig: Results - disc properties}
\end{figure*}

Figure~\ref{fig: Results - disc properties} shows the change of the properties of the disc of Mira~A with pulsation phase, $\phi$, and frequency, $\nu$. 
We analysed three groups of observations within individual pulsation cycles, two of which also cover several bands, thus providing useful constraints on short-timescale behaviour within a single cycle while also allowing comparisons across frequencies. The first consists of observations obtained between September and November 2017, covering phases between $\phi \simeq 0.67$ and $0.82$, around optical minimum light, therefore offering only a relatively short view of the variability within the cycle. These include one epoch in Band~4, four epochs in Band~6, and one epoch in Band~7 (see Table~\ref{tab:observations}). The four Band~6 observations, spanning $\phi\simeq0.68$--$0.72$, show an overall increase in the flux density of the disc of about 25\%. Changes on timescales of days are also apparent, with the measured disc flux density decreasing by approximately 20\% over three days (see Table~\ref{tab:results}). 
A second closely sampled sequence of epochs consists of four Band~6 observations obtained within a few days in September 2021, all near optical maximum light at $\phi\simeq0.08$. Although some small day-to-day variation is present, the measured flux densities are broadly consistent within the adopted uncertainties, which are dominated by the 10\% absolute flux-calibration uncertainty.
Finally, the 2023 observations span approximately $\phi=0.02$--$0.18$ and cover Bands 6 to 9. The two Band~6 epochs in this pulsation cycle are separated by about one month and 5 days ($\Delta\phi\simeq0.1$), during which the measured disc flux density decreases slightly from $\sim$\,117 to 105~mJy. In Band~7, three observations separated by about one week each, spanning $\sim$~15 days in total, show comparatively little change in the measured disc flux density. Between the two Band~9 observations obtained in June 2023 very close to optical maximum ($\phi\simeq0.02$--$0.04$), the flux density of the disc increases by $\sim$~30\% within four days. 

The limited data points across one pulsation phase in each band prevents us from establishing the detailed evolution of the flux density throughout an individual cycle.
Nevertheless, by combining the Band~6 measurements from different pulsation cycles, we can obtain a tentative picture of the flux variation with pulsation phase, as illustrated in Fig.~\ref{fig: Results - disc properties}a. In addition to the three previously discussed pulsation cycles, two further independent measurements at intermediate pulsation phases of $\phi =0.37$ (2025-08-10) and $\phi =0.59$ (2019-06-19) are included when considering the general phase-dependence of the flux density of the fitted disc in Band~6. Taking into account all pulsation cycles together, the highest measured flux densities occur near optical maximum light, around $\phi\simeq0.08$. The observations suggest that the flux density decreases as the star moves away from maximum light, reaching its lowest measured value around $\phi\simeq0.68$, before increasing again at later phases. This behaviour does not appear to be unique to Band~6, although the sampling in Band~7 is even more limited. In Band~7, the flux density measured near minimum light in 2017 ($\phi\simeq0.82$) is approximately 20\% lower than that measured near maximum light in 2023 ($\phi\simeq0.06$). The phases of the highest and lowest measured mm-flux densities broadly coincide with the corresponding maximum and minimum of the optical light curve of Mira~A (see Fig.~\ref{fig: Results - disc properties}a), which contrasts with the behaviour reported at 8~GHz by \citet{RM1997}, where the 8~GHz-minimum flux density was found near $\phi\simeq0.15$. We note, however, that the Band~6 observation closest to optical maximum light corresponds to a phase of $\phi\simeq0.075$. We cannot therefore determine whether the mm-flux density peaks exactly at optical maximum. 
Furthermore, as the measurements shown in Fig.~\ref{fig: Results - disc properties}a combine several pulsation cycles, it is not possible to infer whether the observed differences can be attributed uniquely to pulsation phase-dependent changes. They may also contain cycle-to-cycle variations or shorter-timescale variability occurring within individual cycles. Conversely, Band~6 observations obtained  between two pulsation cycles at nearly the same pulsation phase, $\phi\simeq0.08$, in September 2021 and July 2023 show consistent flux densities within the uncertainties.

\begin{figure}[h]
\centering
\includegraphics[page=1,width=0.45\textwidth]{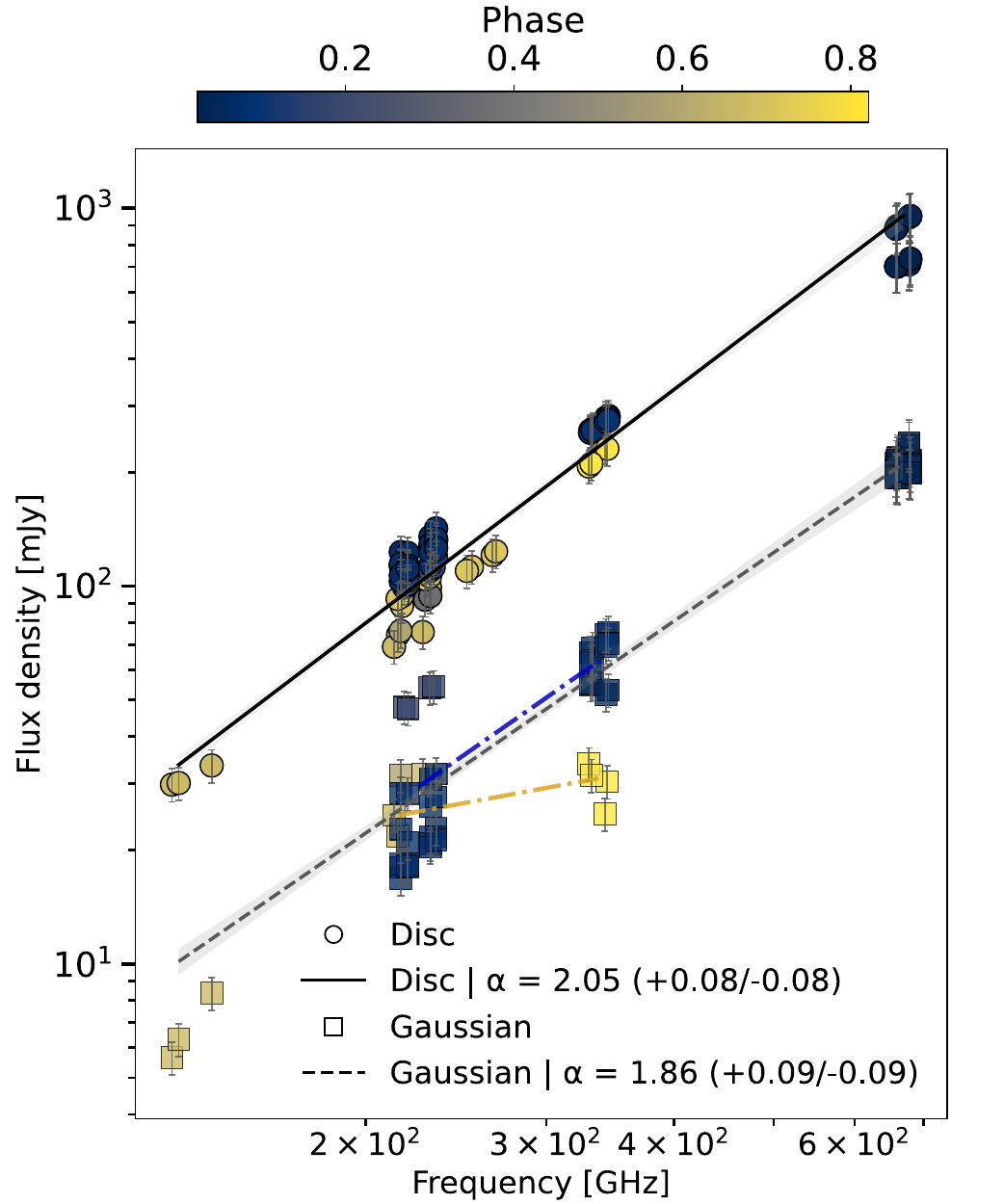}
\caption{Derived flux density and spectral indices ($\alpha$) of the disc (circles) and Gaussian (squares) components as a function of frequency. Each point represents one spw. The solid and dashed lines represent the fit to the flux density-frequency relation for the disc and Gaussian components, respectively. The shaded regions around these lines represent the uncertainties on the fitted relations. The dashed-dotted lines link the median flux of the Gaussian component between Bands~6 and 7 for epochs within a pulsation cycle, in September-November~2017 (yellow, $0.6<\phi<0.9$) and June--July~2023 (blue, $0.0<\phi<0.1$), see Sect~\ref{sec: discussion - stellar activity} for details. }
\label{fig: Results - flux_spId}
\end{figure}

The overall flux density ($F_\nu$) of the disc component increases with observing frequency, yielding an overall spectral index of $\alpha = 2.05 \pm 0.08$ (see Fig.~\ref{fig: Results - flux_spId}), consistent with optically-thick thermal emission ($\alpha = 2$), where $F_\nu \propto \nu^\alpha$. In Band 6, which is the only band for which a statistically significant in-band spectral index could be derived, we obtain a value of $\alpha _\mathrm{B6} = 1.86 ^{+0.61} _{-0.68}$ overall. Considering the variability of the flux density with time and thus separating the epochs in Band~6 into their respective pulsation cycles, we find spectral indices of $\alpha_\mathrm{B6} \simeq 1.84 \pm 0.12$ for epochs near optical maximum ($0.0<\phi<0.2$) in June--August 2023, and of $2.11 \pm 0.16$ for those around minimum light ($0.6<\phi<0.9$) taken in September-November 2017. 
When calculating the spectral index using different subsets of the available bands, we obtain notably different values. The spectral index derived from Bands~4 to 7 is $\alpha _\mathrm{B4-B7} = 2.11\pm0.16$ near optical maximum (June--August 2023), and $\alpha _\mathrm{B4-B7} = 2.05\pm0.18$, near optical minimum (September-November 2017). The spectral indices between Bands~6 and 7 are also consistent with $\alpha = 2 $ for both pulsation phases/cycles when considered separately. On the other hand, a shallower value of $\alpha _\mathrm{B6-B9} = 1.83\pm0.10$ is obtained between Bands~6 and 9, based on observations taken in 2023 within $0.0<\phi<0.2$, due to the even shallower spectral index, $\alpha_\mathrm{B7-B9} = 1.62 \pm 0.19$ between the highest-frequency bands, Bands 7 and 9.  The shallower spectral index toward higher frequencies likely reflects the dependence of the flux density on both the brightness temperature and the apparent size of the disc when the latter is spatially resolved. We therefore calculated the resolved spectral index, defined by 
\begin{equation}
   \alpha_{\rm res}
    =
    \frac{\partial \log T_\mathrm{b}}{\partial \log \nu} + 2
\end{equation}
 \citep{Bojnordi2024}. 
Figure.~\ref{fig: Results - disc properties}d shows that the disc brightness temperature remains approximately constant across the observed frequency range within the relatively large uncertainties. This corresponds to a resolved spectral index of $\alpha_\mathrm{res} = 2.01 \pm 0.15$, consistently seen across epochs, across different subsets of bands, and over the full frequency range. The observed flux density trend with frequency is, in general, consistent with theoretical expectations for radio photospheres \citep{Bojnordi2024} and with previous studies on Mira~A by \citet{Vlemmings2019}, who reported an unresolved spectral index of $2.07 \pm 0.16$ between Bands 3 and 7, and \citet{Matthews2015} who found a flux density behaviour consistent with $\alpha = 1.86$ for fluxes measured at 46, 96, and 229\,GHz. 

The apparent geometric diameter of the stellar disc, defined as $D_\mathrm{geo} = \theta_\mathrm{maj} \sqrt{AR}$, where $\theta_\mathrm{maj}$ denotes the major axis and $AR$ the axis ratio (minor/major axes) of the disc tabulated in Table~\ref{tab:results}, shows a similar tentative phase dependence as the flux density. It ranges between $\sim$\,40 and $\sim$\,48 mas across the observed bands and epochs, with the star appearing largest close to optical maximum light and smallest at minimum light in Band~6 (see Fig.~\ref{fig: Results - disc properties}b). This trend is opposite to that seen at optical/infrared wavelengths, where larger sizes are reported near minimum light \citep[e.g.][]{Wittkowski2018, Jadlovsky2026}. We also find tentative evidence for longer-term size variations when comparing observations obtained at the same pulsation phases ($\phi \simeq 0.08$) but separated by two pulsation cycles. The disc appears to be about 2.8\%  ($\sim$\,1.3\,mas) smaller in July 2023 than in September 2021, significant at the $6\sigma$ level even when accounting for the scatter among the four observations taken in consecutive days in September 2021. However, the limited number of comparable epochs prevents us from drawing robust conclusions about the long-term evolution of the disc size. 
In addition, the multi-cycle size measurements are subject to the same limitations as the flux density, making it difficult to disentangle variations within a pulsation phase and shorter- or longer-timescale variability. 

When considering the geometric diameter as a function of frequency, no clear overall trend is immediately apparent. Since the source size also varies with time, comparisons across frequency are only meaningful when considering observations taken at similar phases within the same pulsation cycle. We therefore analysed the geometric diameter as a function of frequency separately for epochs in June--August 2023, near optical maximum ($0.0<\phi<0.2$), and epochs in September-November 2017, around optical minimum light ($0.6<\phi<0.9$, 2017 cycle). We performed a weighted fit in log–log space, assuming $D_\mathrm{geo} \propto \nu^\eta$, and obtained a slope of $\eta_{(0.0<\phi<0.2)} = - 0.083 ^{+0.004} _{-0.005}$ for the size-frequency relation for observations near maximum light in 2023. As shown in Fig.~\ref{fig: Results - disc properties}e, the disc shows a potential trend of decreasing size with increasing frequency for $0.0<\phi<0.2$, with a clear outlier that lies well above the fitted relation in Band~7. This appears to be mainly due to the lower axis ratios measured in Band~6 ($\sim$~225\,GHz) compared with those in Band~7 ($\sim$~338\,GHz) at similar phases, as shown in Fig.~\ref{fig: Results - disc properties}f. When fitting the major axis instead of the geometric size, we obtain a steeper relation, with $\eta_{(0.0<\phi<0.2,\,\theta\mathrm{maj})}=-0.127^{+0.007}_{-0.008}$ for observations taken in June--July 2023.
On the other hand, around minimum light in 2017 ($0.6<\phi<0.9$), the geometric diameter shows a large scatter, consistent with an approximately flat frequency dependence. In contrast, fitting the major axis for these epochs instead gives $\eta_{(0.6<\phi<0.9,\,\theta\mathrm{maj})}=-0.155^{+0.014}_{-0.013}$ (see Fig.~\ref{fig: Additional plots - axis ratio - disc}b). The difference between the geometric-diameter and major-axis trends is influenced by the particularly low axis ratio ($\sim 0.66$) measured in Band~4 at these phases (see Table~\ref{tab:results}), which biases the geometric size toward smaller values. Our results are in agreement with the slope of $\eta = -0.087\pm0.006$ of the size versus frequency relation found by \citet{Vlemmings2019} between Bands 4 and 6, but for observations close to minimum light ($0.6<\phi<0.9$), where the size of  Mira~A was defined as the mean between the major and minor axes of the fitted disc.

The morphology of the stellar disc also varies with time, as shown by the changes in the axis ratio illustrated in Fig.~\ref{fig: Additional plots - axis ratio - disc}a. Short-term morphological variations on timescales of about a week are seen within a single pulsation cycle in 2023. A large scatter in axis ratio is also observed between epochs separated by only a few days in Band~6, both in October 2017 and September 2021. However, substantial changes in the overall shape of the stellar disc over only a few days are difficult to reconcile with a plausible physical origin and therefore seem unlikely. In general, the disc appears less circular near maximum light, with ellipticities between 15 and 23\% in Band 6. The measured axis ratio approaches unity as the epochs move away from maximum light.  However, the limited number of epochs makes this trend tentative.  In addition, the scatter in the measured axis ratio among the September 2021 observations prevents any robust assessment of long-term morphological changes relative to the 2023-07-13 epoch at a similar phase two cycles later.
 
The increase in flux density and decrease in apparent size with frequency for $0 < \phi < 0.2$ indicate that the (sub)mm emission does not arise from a single blackbody surface, but rather from a stratified atmosphere in which higher frequencies probe deeper layers.
The phase-dependent variations in flux and size likely reflect changes in atmospheric opacity during the pulsation cycle. After maximum light, the propagation of shocks through the atmosphere increases the temperature and the degree of ionisation, thus enhancing the free–free opacity and producing a larger apparent radio photosphere. As the shocks propagate away from the star, the free-free opacity decreases and the gas becomes optically thin first at higher and then at lower frequencies. This leads to a decrease in stellar size with time \citep{Bojnordi2024}. The tentative trend toward a less circular disc near maximum light in Band 6 suggests that the stellar atmosphere displays more features during this phase of the pulsation cycle. This is consistent with the presence of shock-induced layers formed around maximum light. The derived brightness temperatures of the stellar disc span a wide range of values, from $\sim$~1600 to 2200~K across all bands. Given their relatively large uncertainties of $\sim$10\%, these temperatures show no statistically significant trend with pulsation phase, as shown in Fig.\ref{fig: Results - disc properties}c. 

As the observations used in this analysis were obtained over several years with different uv-coverages and noise levels, we investigated whether these differences could reproduce the apparent variability in the fitted source properties using synthetic observations matched to the individual observational setups (see Appendix~\ref{Appendix: subsection - test synthetic data}). Although small systematic biases were identified for some fitted parameters and included in all quoted uncertainties, where relevant (in Table~\ref{tab:results}, Figs.~\ref{fig: Results - disc properties}--\ref{fig: Results - Gaussian properties} and Figs.~\ref{fig: Additional plots - axis ratio - disc}--\ref{fig: Additional plots - Offset}), these effects cannot account for all the overall measured variability. This shows that the observed variations are unlikely to arise from systematic effects and instead trace temporal changes in the source, although some variations in the fitted parameters may reflect complexities not fully captured by our simple geometric models. The PAs are the exception, as they are too poorly constrained to provide a reliable measure of changes in source orientation. The same conclusion applies to the fitted parameters of the Gaussian component discussed in Sect.~\ref{sec: discussion - stellar activity}. 

\subsection{Surface activity}
\label{sec: discussion - stellar activity}
\begin{figure*}[h]
\centering
\includegraphics[page=1,width=1\textwidth]{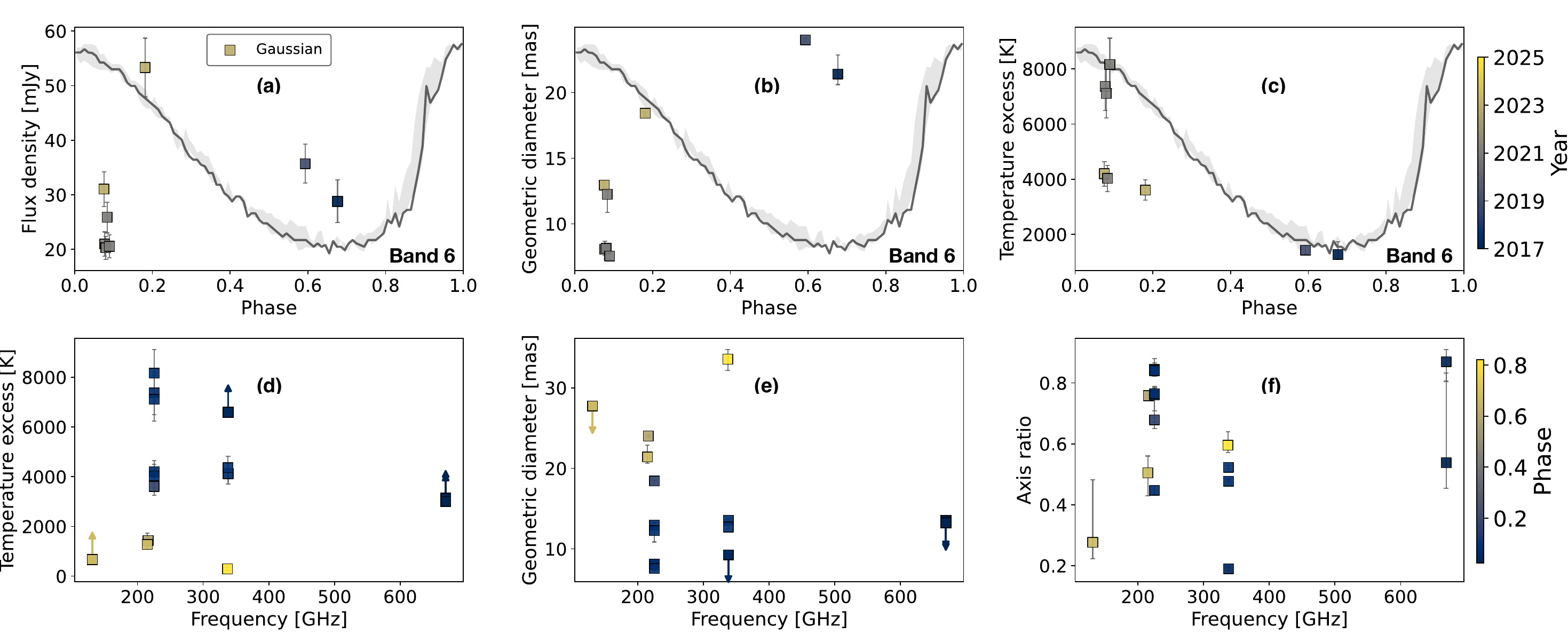}
\caption{Parameters of the fitted Gaussian component: \textbf{(a)} Flux density of the fitted Gaussian as a function of pulsation phase in Band 6 scaled at a common frequency of 231~GHz, using a spectral index of $\alpha=1.86$ (see text for details); \textbf{(b)} Geometric diameter of the Gaussian in Band 6 as a function of phase; \textbf{(c)} Excess brightness temperature in Band 6 as a function of phase; \textbf{(d)}~Excess brightness temperature versus frequency;  \textbf{(e)} Geometric diameter as a function of frequency;  \textbf{(f)} Minor-to-Major axis ratio of the Gaussian component versus frequency. In all panels, one square symbol represents one observing epoch. The solid line (upper panels) represents the folded visual light curve of Mira A, as described in Fig.~\ref{fig: Results - disc properties}.}
\label{fig: Results - Gaussian properties}
\end{figure*}

Similar to the stellar disc, the flux density of the Gaussian component varies with time.
As shown in Fig.~\ref{fig: Results - Gaussian properties}a, the measured Band~6 flux densities are lowest close to maximum light ($\phi\simeq0.08$), where the Gaussian flux density is approximately 15 to 27\% of that of the stellar disc. Near maximum light, the flux density of the Gaussian component shows a notable scatter, even after scaling the measurements to a common frequency, with differences exceeding that expected from absolute flux-calibration uncertainties, potentially indicating short- and/or long-term variability.
On the one hand, we observe evidence of variations on timescales of days within a single pulsation cycle. For instance, on 2021-09-21 ($\phi\simeq0.08$), the Gaussian flux density is about 22\% higher than on the preceding day and two days later (see Table~\ref{tab:results}). On the other hand, a comparison of the median flux density measured in September 2021 with the flux density obtained on 2023-07-13 at the same pulsation phase but two cycles later shows a flux difference of $\sim$~50\%. However, given the short-time variability and the lack of better sampling within and between each cycle, we cannot determine whether the difference between 2021 and 2023 reflects a longer-term change between pulsation cycles or short-term variability within a cycle. Furthermore, the flux density measured on 2021-09-21, which is higher than that measured on the surrounding days, is consistent within the uncertainties with the flux measured in 2023.

Moving slightly away from maximum light during the 2023 pulsation cycle, the flux density of the fitted Gaussian increases by approximately 70\% between July and August 2023 ($\phi\simeq0.08{-}0.18$), reaching its highest value at $\phi\simeq0.18$, where it corresponds to about half of the stellar-disc flux. Near minimum light ($\phi\simeq0.6$), the Gaussian shows flux values of approximately 30 to 40\% of the disc flux, based on measurements obtained in different years and pulsation cycles. In Band~7, a broadly similar behaviour is observed, where the highest flux density is measured at $\phi\simeq0.10$ rather than closer to maximum light, and the lowest at $\phi\simeq0.82$, with a Gaussian-to-disc flux ratio of about 0.15 for the latter. However, the relatively large uncertainties and limited temporal sampling in Band~7, together with the fact that the observations span different pulsation cycles but cover different pulsation phases, prevent us from drawing any firm conclusions. Overall, the available data provide evidence for variations on timescales of days to weeks within individual cycles, but the sampling is insufficient to determine when the Gaussian flux reaches its absolute maximum value or to constrain variability on longer timescales, either over several months within a cycle or between different pulsation cycles.

The flux density of the Gaussian component shows an overall increase with observing frequency, corresponding to a global spectral index of $\alpha_\mathrm{Gauss} = 1.86 \pm 0.09$, as shown in Fig.~\ref{fig: Results - flux_spId}. However, given the complexity of the variability of the Gaussian component, we can only investigate a meaningful spectral index from observations acquired within the same pulsation cycle and at similar pulsation phases. Since the flux density of the Gaussian already differs substantially, by about a factor of 2, between $\phi \simeq 0.08$ and $\phi \simeq 0.18$ in Band~6, we limit the derivation of the spectral index close to maximum light to epochs within $0.0 <\phi \leq 0.1$ (June--July 2023), for which we obtain an overall spectral index of $\alpha_\mathrm{Gauss} = 1.77 \pm0.13$. For epochs around minimum light ($0.6 <\phi < 0.9$, September-November 2017), we find $\alpha_\mathrm{Gauss} = 1.55 ^{+0.44}_{-0.39}$. Interestingly, the spectral index between Bands~6 and 7 is $\alpha_\mathrm{Gauss} = 1.87 ^{+0.30}_{-0.28}$ for $0.0 <\phi \leq 0.1$, while it is much flatter near minimum light, with a shallower slope (see Fig.~\ref{fig: Results - flux_spId}) that is not sufficiently well constrained to be robustly quantified.

Overall, the smallest measured sizes of the fitted Gaussian component are observed near maximum light, while larger sizes are measured at later pulsation phases, as shown in Fig.~\ref{fig: Results - Gaussian properties}b for Band~6 observations. When all pulsation cycles are considered, the geometric diameter of the fitted Gaussian ranges from approximately 7.5 to 13~mas around maximum light when the component is resolved. An exception is the Band~7 observation from 2023-07-06, for which the minor axis of the compact component remains unresolved.
We observe changes in the diameter of the Gaussian component on timescales of days within a single pulsation cycle, with one epoch showing a larger size by nearly 60\% than the surrounding days. Interestingly, this larger value is comparable to that measured at approximately the same pulsation phase ($\phi\simeq0.08$) two pulsation cycles later, differing by only about 6\%, making the nature and timescale of the variation even more ambiguous. On the other hand, the Gaussian component appears largest near minimum light, where its size exceeds 50\% of the stellar-disc geometric diameter. For the Band~4 2017-09-21 and Band~6 2017-09-22 epochs ($\phi\simeq0.67$), the Gaussian component extends partially beyond the stellar disc. In these cases, the fitted Gaussian may represent a distinct structure on the stellar surface, but it could also instead reflect a departure from a symmetric stellar disc. We note, however, that these two epochs have among the largest size uncertainties in the sample and correspond to the largest synthesised beams (see Tables~\ref{tab:observations} and \ref{tab:results}). For the Band~4 epoch, the limited angular resolution leaves the Gaussian component unresolved. In general, the same sampling limitations discussed for the Gaussian flux density also apply here, making the observed size variation with time difficult to interpret robustly.

Within a single pulsation cycle, the size of the Gaussian component shows no clear frequency dependence (see Fig.~\ref{fig: Results - Gaussian properties}e). The morphology of the Gaussian component also shows no strong overall dependence on frequency, although the structure appears roughly more circular in Band~6 than in Band~7 at similar pulsation phases, as shown in Fig.~\ref{fig: Results - Gaussian properties}f. In both Bands~6 and 7, the axis ratio increases by $\geq$~30\% between July and August 2023 ($\phi\simeq0.05{-}0.2$), hinting at the structure becoming less elongated as the star moves away from maximum light in that pulsation cycle. However, the compact nature of the component, the intricacies of the stellar surface, and possible degeneracies in the fitted models make the significance of this tentative trend difficult to assess.

The Gaussian component can reach brightness temperatures a few times higher than those of the surrounding stellar disc, as seen in Figs.~\ref{fig: Results - Gaussian properties}c and~\ref{fig: Results - Gaussian properties}d, depending on phase and frequency. In Band~6, the excess brightness temperature relative to the disc temperature near maximum light ($\phi \simeq 0.08$) is up to $\sim$\,$8200$\,K in September 2021. It is determined to be $\Delta T \simeq 4000$\,K in July 2023, at a similar phase but two pulsation cycles later. For the 2023 pulsation cycle, for which we have observations in Bands~6, 7, and 9 at a similar phase near maximum light, the properties derived for the Gaussian component in Bands~6 and 7 are similar, with excess temperatures around $\sim$\,4000\,K. In Band~9, the fitted Gaussian is larger and colder, with a temperature excess of around $2900$\,K. However, given the complexity of the data and model degeneracy (see Appendix~\ref{Appendix: subsection - degeneracy}), the true brightness temperature could be significantly higher. For example, if the excess emission came instead from multiple unresolved components, the brightness temperature of the spots on the continuum surface in Band~9 would exceed $7700$\,K. In contrast, around minimum light, the corresponding brightness temperature excess is less significant, as the measured structures are larger. Hotspots are therefore a prominent feature of the (sub)mm stellar surface around optical maximum light. We note that the brightness temperatures shown in Figs.~\ref{fig: Band 9} and ~\ref{fig: Continuum}--~\ref{fig: Continuum 2} are beam-averaged, such that compact emission on angular scales smaller than the synthesised beam is diluted and can therefore, in some epochs, appear substantially cooler than the brightness temperatures inferred from the visibility-fitted component sizes.

When the same hotspot can be confidently traced across multiple epochs within a pulsation cycle, it shows small positional displacements on short timescales (see Fig.~\ref{fig: Additional plots - Offset}). In Band~6 at $\phi \simeq 0.08$, the radial offset increases by $\sim1.25$~mas within five days in September 2021. In July 2023, the Gaussian component moves closer to the stellar centre by $\sim3.5$~mas over about one month ($\Delta \phi \simeq 0.11$). In Band~7, the radial offset increases from $\sim4$ to $\sim12$~mas within one week and shows little change thereafter. However, it is unclear whether the 2023-07-06 best-fit component traces the same structure seen in the following weeks. Its smaller size and more elongated shape may instead reflect model degeneracy. In Band~9 the radial offset decreases slightly by $\sim$\,$1.7$~mas within a few days. During the June–July 2023 observations, obtained within $\sim$\,$15$ days, the hotspot lies at projected distances from the centre corresponding to $\sim$\,$68$, $55$, and $49$\% of the stellar radius in Bands~6, 7, and 9, respectively, suggesting that the emission originates closer to the stellar centre at higher frequencies, although short-term variability may also contribute. Based on the observed displacements in Bands~6 and~7 within a week, we infer projected velocities of at least $39$\,km s$^{-1}$, much larger than expected from the slow rotation of an AGB star \citep{Vlemmings2018}. On longer timescales, the positions of the fitted Gaussian components differ between pulsation cycles. The Gaussian component appears in the NW quadrant of the stellar disc in the September 2021 epochs, while two pulsation cycles later, at a similar pulsation phase, the fitted component is located in the SW quadrant in the June--July 2023 observations (see Figs.~\ref{fig: Continuum} and \ref{fig: Additional plots - Offset}), with radial offsets of $\sim7$ and $\sim15.5$~mas, respectively. 

The apparent changes in the position of the compact component may in principle be affected by residual calibration and other systematic effects. We assessed the contribution from differences in uv-coverage, noise level, and model recovery using validation tests on synthetic observations, as described in Appendix~\ref{Appendix: subsection - test synthetic data}. These tests indicate systematic uncertainties on the radial position of the Gaussian component of up to $\sim$0.08, 0.04, and 0.01\,mas in Bands~6, 7, and 9, respectively (see Table~\ref{tab:real_synthetic_rms_unclustered}). The observed changes in the fitted position of the compact component are larger than these systematic contributions and, therefore, cannot be explained solely by the differences in uv-coverage and noise properties. At high angular resolution, the absolute astrometric uncertainty of ALMA observations can be on the order of 10\% of the synthesised beam, corresponding to approximately 1.1--1.9~mas for the epochs near maximum light in September 2021 and June--August 2023. Although the Gaussian position is measured relative to the fitted disc centre, residual epoch-dependent phase errors may still affect the relative position of the Gaussian component and contribute to the apparent epoch-to-epoch displacements. Therefore, the physical significance of displacements of the order of 1--2\,mas, as measured here for the Gaussian component on timescales of days, remains uncertain, whereas the larger measured displacements observed over longer timescales are less likely to be explained by these calibration uncertainties alone.

\subsection{Interpretations}

Surface inhomogeneities on the (sub)mm photosphere of Mira~A and other AGB stars have previously been attributed to convective cells \citep[e.g.][]{Matthews2015, Matthews2018, Vlemmings2024}. Such features are typically large, grow with pulsation phase \citep{Rosales2024}, and exhibit low brightness contrasts with temperature enhancements of only $700{-}1500$~K in Band~7 \citep{Vlemmings2024}. In contrast, the hotspots detected here around maximum light show much larger brightness temperature excesses, from $\sim$\,2900~K to several thousand kelvin above the stellar disc for $0.0 <\phi < 0.2$ in all bands, reaching up to an excesses of $8200$\,K in Band~6 for a size of $\sim 8 \times 7$~mas. These compact hotspots observed near maximum light in Bands 6 and 7 are unlikely to correspond to the convective cells predicted by current 3D AGB atmosphere models. Their origin, therefore, remains uncertain. Strong shocks have been proposed as a possible explanation \citep{Vlemmings2015, Vlemmings2017}, but current AGB atmosphere simulations show that rapid radiative cooling limits the temperature enhancements produced by pulsation- and convection-driven shocks \citep{Freytag2017}. Magnetic activity may therefore play an additional role. Soft X-ray outbursts observed from Mira~A have been attributed to magnetic flares with lifetimes of weeks \citep{Karovska2005}, comparable to the persistence timescales of the hotspots detected here. Recent maser observations tracing a magnetic eruption event around Mira~A \citep{Vlemmings2026} suggest that magnetic fields may contribute significantly to the atmospheric dynamics and could be related to the observed hotspots. A binary origin for these hotspots is unlikely, as the Mira~AB system is weakly interacting at a projected separation of $\sim40$\,au ($\gtrsim20$ stellar radii). Furthermore, the hotspots are not preferentially located on the side of Mira~A facing Mira~B, pointing to an origin intrinsic to  Mira~A.

The detection of the hotspot in multiple ALMA bands in June–July 2023 suggests that the structure likely spans several layers of the radio photosphere. The spot appears at different projected radial distances from the stellar centre in Bands~6, 7, and 9. The projected position of the spot closer to the centre of the stellar disc at higher frequencies (where the star is smaller) implies that the underlying feature is an elongated structure extending across these three continuum surfaces, possibly with a notable tangential component.

It is unclear if these spots are part of the ejection mechanism that creates the powerful ejections reported by \cite{Khouri2026}. Assuming the ejections observed by \citet{Khouri2026} originate from a single event, their kinetic energy is $\sim$\,$2 \times 10^{41}$\,erg. In comparison, the thermal energy of the hotspot observed in September 2021, with a characteristic size of $\sim$~0.75\,au, a temperature excess of $\sim8200$\,K, and density $n \sim 10^{12}$\,cm$^{-3}$ \citep{Vlemmings2019, Bojnordi2024}, is $>1.0 \times 10^{40}$\,erg. A similar energy is obtained for the larger hotspot observed in July 2023 ($\Delta T \simeq 4200$\,K, size $\sim1.9 \times 0.9$\,au, with a characteristic vertical scale of $\sim0.8$\,au, assuming that the structure spans the layers probed by Bands~6 and~9). On the other hand, the kinetic energy of a typical AGB wind for Mira A, with a terminal velocity of $8$\,km\,s$^{-1}$, a mass-loss rate of $1.1 \times 10^{-7}\,M_\odot\,\mathrm{yr}^{-1}$ \citep{DeBeck2010}, and assuming a total envelope mass of $3 \times 10^{-3}\,M_\odot$ \citep{Martin2007}, is $\sim2 \times 10^{42}$\,erg. These values suggest that hotspots represent an important energy reservoir in the stellar atmosphere of Mira~A and could play a role in episodic mass loss events.

The absence of prominent hotspots away from maximum light suggests that these structures are likely short-lived features that appear and dissipate over the course of a pulsation cycle, rather than representing a single long-lived spot. The variation of the spectral index between Bands 6 and 7 from $\sim1.85$ close to maximum light in 2023 to values approaching zero around minimum light in late 2017 is consistent with this interpretation, as it can reflect a transition from optically thick to optically thin free–free emission. This behaviour indicates that energetic events of this type may recur across pulsation cycles and could represent a non-negligible contribution to the energy budget associated with the mass-loss process.
However, the limited temporal sampling of current high-angular-resolution observations prevents us from systematically tracing the evolution of these structures. More regular access to ALMA's extended configurations, as also advocated by \citet{Velilla-Prieto2026}, would enable systematic monitoring over consecutive pulsation cycles, ultimately helping to establish the physical origin of these hotspots and their potential role in shaping the extended atmospheres and mass loss of AGB stars. 

\begin{acknowledgements}
We thank the anonymous referee for their constructive comments, which helped improve the manuscript. MA, BB, WV, TK, and LP acknowledge support from the
Olle Engkvist Foundation under project 229-0368. JA acknowledge support from project CRISPNESS, grant PID2023-146056NB-C21, funded by MICIU/AEI/10.13039/501100011033 and by ERDF/EU. KTW acknowledges support from the European Research Council (ERC) under the European Union's Horizon 2020 research and innovation programme (Grant agreement no. 883867, project EXWINGS) and support from the Knut and Alice Wallenberg Foundation under grant no. KAW 2023.0314. This paper makes use of the following ALMA data:
ADS/JAO.ALMA\#\allowbreak 2022.1.01071.S,
ADS/JAO.ALMA\#\allowbreak 2022.1.00817.S,
ADS/JAO.ALMA\#\allowbreak 2017.1.00191.S,
ADS/JAO.ALMA\#\allowbreak 2019.1.01072.S,
ADS/JAO.ALMA\#\allowbreak 2024.1.01776.S,
ADS/JAO.ALMA\#\allowbreak 2018.1.00551.S,
ADS/JAO.ALMA\#\allowbreak 2017.1.00393.S, and
ADS/JAO.ALMA\#\allowbreak 2016.1.00004.S.
ALMA is a partnership of ESO (representing its member states), NSF (USA), and 
NINS (Japan), together with NRC (Canada), MOST and ASIAA (Taiwan), and KASI 
(Republic of Korea), in cooperation with the Republic of Chile. The Joint ALMA 
Observatory is operated by ESO, AUI/NRAO, and NAOJ. We gratefully acknowledge the contributions of the AAVSO observer community, whose photometric data and metadata resources were used in this study and made available through the AAVSO's scientific archives.
\end{acknowledgements}

\bibliographystyle{aa} 
\bibliography{biblio} 

\begin{appendix}
\clearpage
\onecolumn
\section{Observations and images}
\label{Appendix: observations}
\begin{table}[H]
\centering
\footnotesize
\caption{Details of observations.}
\label{tab:observations}
\setlength{\tabcolsep}{2.75pt}
\renewcommand{\arraystretch}{1.15}
\begin{tabular}{cccccl}
\hline
\toprule
Date & Central Freq. & Superuniform & Uniform & nspws & ALMA project \\
 &  & $\theta_\mathrm{beam}$, PA, rms & $\theta_\mathrm{beam}$, PA, rms &  & \\
& [GHz] & [mas $\times$ mas], [$^\circ$], [mJy\,beam$^{-1}$] & [mas $\times$ mas], [$^\circ$], [mJy\,beam$^{-1}$] & & \\
\midrule
\multicolumn{6}{c}{Band 4} \\
\hline
2017-09-21 & 131.30 & 40.89 $\times$ 26.93, 35.64, 0.12 & 46.98 $\times$ 36.53, 40.22, 0.30 & 3 & 2016.1.00004.S \\
\hline
\multicolumn{6}{c}{Band 6} \\
\hline
2017-09-22 & 215.02 & 30.95 $\times$ 18.90, 43.38, 0.30 & 38.87 $\times$ 23.99, 44.42, 0.44 & 3 & 2016.1.00004.S \\
2017-10-04 & 259.86 & 15.67 $\times$ 13.27, 78.57, 0.13 & 17.92 $\times$ 13.73, 88.80, 0.15 & 4 & 2017.1.00393.S \\
2017-10-05 & 222.92 & 16.74 $\times$ 14.18, -86.80, 0.08 & 19.83 $\times$ 14.72, -86.30, 0.13 & 4 & 2017.1.00393.S \\
2017-10-06 & 223.92 & 17.20 $\times$ 14.22, 77.86, 0.12 & 18.97 $\times$ 15.19, -85.24, 0.19 & 2 & 2017.1.00393.S \\
2019-06-19 & 216.18 & 17.32 $\times$ 15.67, 56.67, 0.08 & 19.05 $\times$ 17.07, -59.51, 0.39 & 1 & 2018.1.00551.S \\
2021-09-19 & 225.48 & 15.05 $\times$ 14.21, 51.93, 0.10 & 16.74 $\times$ 15.08, -77.09, 0.14 & 4 & 2019.1.01072.S \\
2021-09-20 & 225.48 & 15.35 $\times$ 13.99, 58.39, 0.10 & 16.13 $\times$ 16.02, 48.45, 0.16 & 4 & 2019.1.01072.S \\
2021-09-21 & 225.48 & 18.82 $\times$ 13.90, 30.80, 0.17 & 19.54 $\times$ 16.87, 32.71, 0.24 & 4 & 2019.1.01072.S \\
2021-09-23 & 225.48 & 17.33 $\times$ 14.20, 59.25, 0.10 & 17.57 $\times$ 16.81, -89.08, 0.12 & 4 & 2019.1.01072.S \\
2023-07-13 & 225.49 & 19.22 $\times$ 13.95, 31.72, 0.12 & 21.22 $\times$ 16.74, 47.46, 0.12 & 4 & 2022.1.00817.S \\
2023-08-17 & 225.42 & 17.67 $\times$ 14.64, 76.97, 0.12 & 21.47 $\times$ 15.80, 81.65, 0.11 & 4 & 2022.1.01071.S \\
2025-08-10 & 229.83 & 15.70 $\times$ 14.03, 59.34, 0.18 & 18.88 $\times$ 15.37, 73.96, 0.22 & 2 & 2024.1.01776.S \\
\hline
\multicolumn{6}{c}{Band 7} \\
\hline
2017-11-09 & 337.10 & 23.50 $\times$ 13.06, 54.13, 0.22 & 28.05 $\times$ 13.99, 57.62, 0.25 & 4 & 2017.1.00191.S \\
2023-07-06 & 338.06 & 17.12 $\times$ 13.41, -23.37, 0.63 & 20.25 $\times$ 14.45, -47.91, 0.62 & 4 & 2022.1.01071.S \\
2023-07-14 & 338.06 & 13.75 $\times$ 10.21, 32.06, 0.31 & 14.89 $\times$ 11.60, 44.02, 0.26 & 4 & 2022.1.01071.S \\
2023-07-21 & 337.95 & 11.74 $\times$ 10.25, 56.22, 0.30 & 13.44 $\times$ 11.00, 84.01, 0.35 & 4 & 2022.1.00817.S \\
\hline
\multicolumn{6}{c}{Band 9} \\
\hline
2023-06-26 & 669.24 & 13.43 $\times$ 10.38, -55.57, 1.50 & 15.72 $\times$ 11.06, -63.20, 4.51 & 8 & 2022.1.01071.S \\
2023-06-30 & 669.25 & 14.66 $\times$ 12.28, 50.65, 1.51 & 16.59 $\times$ 12.63, 59.06, 2.24 & 8 & 2022.1.01071.S \\
\bottomrule
\end{tabular}
\tablefoot{Central Freq. represents the frequency at the midpoint of the spws; nspws gives the number of considered spws. The beam size ($\theta_\mathrm{beam}$), PA, and rms are reported for the superuniform- and uniform-weighted images. }
\\[0.5ex]
\end{table}

\begin{figure}[hbt]
\centering
\includegraphics[page=1,width=0.4\textwidth]{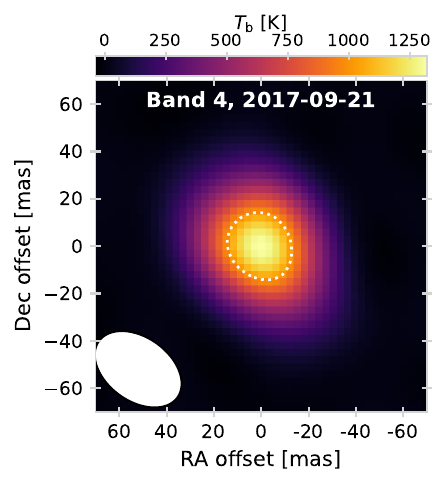}
\caption{Continuum map of Mira~A in ALMA~Band~4 generated using superuniform weighting. The white-dotted contours represent emission at 1000\,K. The rms noise level is $\sim7.8$~K. The white ellipse at the bottom-left corner of the image represents the synthesised beam (see Table~\ref{tab:observations} for beam size values).}
\label{fig: Continuum 2}
\end{figure}

\begin{figure}[hbt]
\centering
\includegraphics[page=1,width=0.95\textwidth]{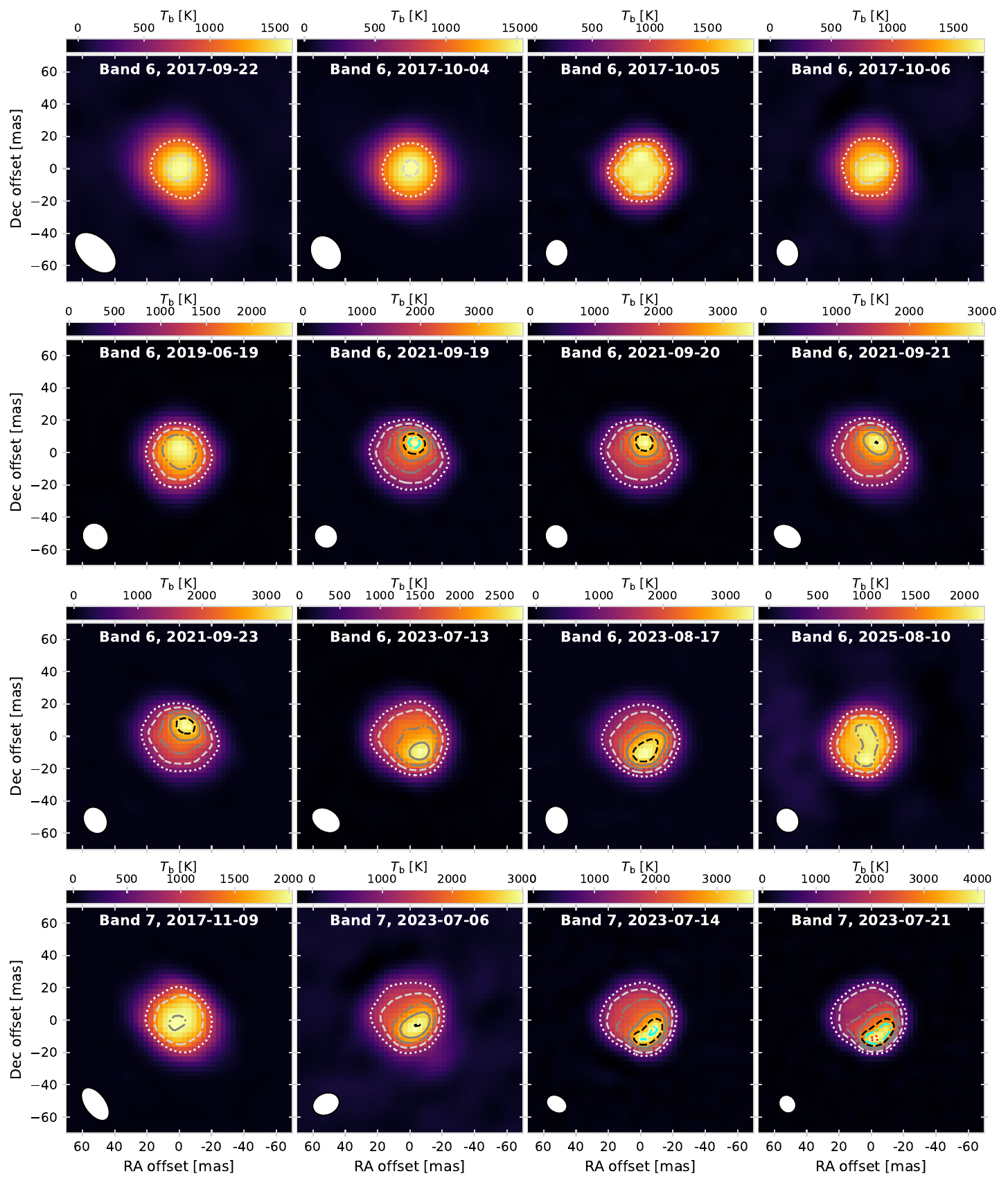}
\caption{Continuum maps of Mira~A in ALMA Bands 6 and 7 generated using superuniform weighting. The contours represent emission at 1000\,K (white dotted), 1500\,K (light gray dashed), 2000\,K (dark gray dot-dashed), 2500\,K (gray solid), 3000\,K (black dashed), 3500\,K (cyan), and 4000\,K (red). The white ellipse at the bottom-left corner of each image represents the synthesised beam (see Table~\ref{tab:observations} for beam size values and the corresponding rms noise levels). The brightness temperatures shown here are beam-averaged. High temperatures inferred for compact fitted components smaller than the synthesised beam are diluted by the beam and are therefore not directly visible in these images.}
\label{fig: Continuum}
\end{figure}

\FloatBarrier
\section{Additional plots and table}
\begin{sidewaystable}[ht]
\centering
\fontsize{8.55}{9.65}\selectfont
\setlength{\tabcolsep}{1.22pt}
\renewcommand{\arraystretch}{1.12}
\caption{Summary of the fitted component properties.}
\begin{tabular}{l c c c c c c c c c c c c}
\hline
\toprule
Epoch & $\nu$ & $\mathrm{S/N}_\star$ & $s_{\rm lim}$ & Comp. & $\phi$ & $F_\nu$ & $\theta_\mathrm{maj}$ & $D_\mathrm{geo}$ & Axis ratio & PA$^\ddagger$ & $T_\mathrm{b}$ & Excess $T_\mathrm{b}$ \\
 & [GHz] & & [mas $\times$ mas] & & & [mJy] & [mas] & [mas] & (\%) & [$^\circ$] & [K] & [K] \\
\midrule
\multicolumn{13}{c}{Band 4} \\
\hline
2017-09-21 & 131.30 & 21.0 & $12.08 \times 9.39$ & D & 0.673 & $30.12\,(-3.01,+3.39)$ & $48.79\,(-0.60,+0.33)$ & $39.54\,(-0.47,+0.38)^{\dagger}$ & $65.68\,(-0.94,+1.02)^{\dagger}$ & $-54.6\,(-1.1,+47.7)$ & $1966\,(-213,+209)$ &  \\
 &  &  &  & G &  & $6.31\,(-0.66,+2.06)$ & $<31.46$$^{*}$ & $<27.74$$^{*}$ & $27.63\,(-5.32,+20.60)$ & $40.7\,(-21.9,+60.5)$ &  & $>668$ \\
\hline
\multicolumn{13}{c}{Band 6} \\
\hline
2017-09-22 & 215.02 & 55.9 & $6.12 \times 3.78$ & D & 0.676 & $74.29\,(-7.43,+7.43)$ & $46.25\,(-0.28,+0.47)^{\dagger}$ & $39.88\,(-0.59,+1.24)^{\dagger}$ & $73.01\,(-1.15,+6.03)$ & $-55.5\,(-1.3,+4.0)$ & $1677\,(-175,+174)$ &  \\
 &  &  &  & G$^{**}$ &  & $24.75\,(-2.91,+6.96)$ & $30.53\,(-1.25,+2.12)$ & $21.39\,(-0.77,+1.46)$ & $50.48\,(-7.48,+5.53)$ & $5.0\,(-12.6,+1.9)$ &  & $1269\,(-195,+461)$ \\
2017-10-04 & 259.86 & -- & -- & D & 0.713 & $116.58\,(-11.66,+11.66)$ & $43.90\,(-0.36,+0.49)$ & $43.67\,(-0.37,+0.56)$ & $99.07\,(-0.40,+0.24)$ & $-22.4\,(-12.6,+6.8)$ & $1592\,(-163,+164)$ &  \\
2017-10-05 & 222.92 & -- & -- & D & 0.716 & $99.20\,(-9.92,+9.92)$ & $43.45\,(-0.07,+0.22)$ & $42.59\,(-0.14,+0.20)^{\dagger}$ & $96.09\,(-0.53,+0.28)^{\dagger}$ & $-13.4\,(-0.8,+1.2)$ & $1930\,(-186,+203)$ &  \\
2017-10-06 & 223.92 & -- & -- & D & 0.719 & $93.81\,(-9.38,+9.38)$ & $48.76\,(-0.62,+0.62)^{\dagger}$ & $45.02\,(-0.38,+0.38)^{\dagger}$ & $85.27\,(-0.77,+0.77)^{\dagger}$ & $3.9\,(-0.5,+0.5)$ & $1624\,(-167,+158)$ &  \\
2019-06-19 & 216.18 & 80.3 & $2.50 \times 2.24$ & D & 0.593 & $76.07\,(-7.61,+7.61)$ & $44.26\,(-0.08,+0.07)^{\dagger}$ & $42.71\,(-0.09,+0.09)^{\dagger}$ & $93.10\,(-0.21,+0.21)$ & $-9.2\,(-0.9,+0.9)$ & $1567\,(-144,+159)$ &  \\
 &  &  &  & G &  & $31.56\,(-3.16,+3.16)$ & $27.58\,(-0.10,+0.10)$ & $24.01\,(-0.14,+0.13)^{\dagger}$ & $75.81\,(-0.63,+0.63)^{\dagger}$ & $69.4\,(-0.5,+0.5)$ &  & $1429\,(-148,+143)$ \\
2021-09-19 & 225.48 & 139.6 & $1.67 \times 1.51$ & D & 0.077 & $129.15\,(-12.91,+12.98)$ & $51.84\,(-0.44,+0.16)^{\dagger}$ & $46.79\,(-0.06,+0.64)^{\dagger}$ & $82.36\,(-1.40,+0.89)$ & $65.4\,(-1.1,+1.6)$ & $2045\,(-206,+219)$ &  \\
 &  &  &  & G &  & $19.91\,(-1.99,+2.94)$ & $9.28\,(-0.41,+0.57)$ & $8.07\,(-0.50,+0.59)^{\dagger}$ & $76.05\,(-5.32,+2.48)^{\dagger}$ & $67.4\,(-8.8,+12.5)$ &  & $7323\,(-840,+910)$ \\
2021-09-20 & 225.48 & 118.9 & $1.75 \times 1.73$ & D & 0.080 & $119.48\,(-11.95,+11.99)$ & $52.06\,(-0.50,+0.59)^{\dagger}$ & $46.92\,(-0.15,+1.02)^{\dagger}$ & $81.99\,(-1.22,+0.97)$ & $65.7\,(-1.8,+1.0)$ & $1870\,(-192,+199)$ &  \\
 &  &  &  & G$^{**}$ &  & $19.41\,(-2.60,+2.32)$ & $9.28\,(-0.30,+0.35)$ & $8.13\,(-0.77,+0.24)^{\dagger}$ & $76.49\,(-11.43,+2.45)^{\dagger}$ & $65.0\,(-2.6,+36.5)$ &  & $7130\,(-872,+949)$ \\
2021-09-21 & 225.48 & 102.9 & $2.27 \times 1.96$ & D & 0.083 & $108.51\,(-10.85,+10.85)$ & $53.51\,(-0.32,+0.49)$ & $46.87\,(-0.13,+0.89)$ & $77.18\,(-0.84,+1.06)$ & $65.7\,(-1.6,+1.8)$ & $1710\,(-173,+177)$ &  \\
 &  &  &  & G &  & $24.44\,(-3.48,+3.22)$ & $13.34\,(-1.36,+0.22)$ & $12.26\,(-1.39,+0.27)^{\dagger}$ & $84.40\,(-2.79,+2.67)^{\dagger}$ & $86.7\,(-9.2,+5.4)$ &  & $4023\,(-462,+508)$ \\
2021-09-23 & 225.48 & 161.3 & $1.63 \times 1.56$ & D & 0.089 & $121.56\,(-12.16,+12.16)$ & $51.19\,(-0.47,+0.31)$ & $46.49\,(-0.06,+0.58)$ & $83.54\,(-2.12,+0.25)$ & $64.4\,(-0.5,+2.8)$ & $1949\,(-207,+201)$ &  \\
 &  &  &  & G &  & $19.39\,(-1.94,+1.94)$ & $8.14\,(-0.18,+0.47)$ & $7.56\,(-0.26,+0.34)$ & $84.05\,(-1.40,+4.02)$ & $80.1\,(-31.0,+16.6)$ &  & $8179\,(-915,+894)$ \\
2023-07-13 & 225.49 & 254.7 & $1.57 \times 1.24$ & D & 0.075 & $116.92\,(-11.69,+11.69)$ & $51.11\,(-0.27,+0.31)^{\dagger}$ & $45.45\,(-0.06,+0.11)^{\dagger}$ & $79.25\,(-1.06,+0.63)^{\dagger}$ & $-68.6\,(-0.2,+0.7)$ & $1954\,(-205,+199)$ &  \\
 &  &  &  & G &  & $29.55\,(-2.96,+2.96)$ & $19.39\,(-0.31,+0.60)$ & $12.96\,(-0.15,+0.31)^{\dagger}$ & $44.75\,(-1.57,+1.63)^{\dagger}$ & $-55.8\,(-0.8,+0.7)$ &  & $4212\,(-425,+426)$ \\
2023-08-17 & 225.42 & 456.4 & $1.18 \times 0.87$ & D & 0.181 & $105.44\,(-10.54,+10.54)$ & $49.16\,(-0.28,+0.19)$ & $45.18\,(-0.13,+0.22)$ & $84.60\,(-0.27,+0.36)$ & $-75.5\,(-0.5,+0.6)$ & $1784\,(-187,+185)$ &  \\
 &  &  &  & G &  & $50.77\,(-5.08,+5.08)$ & $22.30\,(-0.14,+0.12)$ & $18.42\,(-0.23,+0.11)^{\dagger}$ & $67.90\,(-0.94,+0.93)^{\dagger}$ & $-44.4\,(-1.1,+1.1)$ &  & $3606\,(-385,+371)$ \\
2025-08-10 & 229.83 & -- & -- & D & 0.366 & $93.08\,(-9.31,+9.31)$ & $47.35\,(-0.05,+0.05)^{\dagger}$ & $44.11\,(-0.04,+0.04)^{\dagger}$ & $86.79\,(-0.13,+0.13)^{\dagger}$ & $-4.6\,(-0.4,+0.4)$ & $1595\,(-159,+167)$ &  \\
\hline
\multicolumn{13}{c}{Band 7} \\
\hline
2017-11-09 & 337.10 & 123.5 & $2.97 \times 1.48$ & D & 0.821 & $220.95\,(-22.09,+22.09)$ & $42.15\,(-0.52,+0.17)$ & $41.24\,(-0.28,+0.22)^{\dagger}$ & $96.05\,(-0.74,+0.77)^{\dagger}$ & $-15.8\,(-2.4,+1.2)$ & $2003\,(-212,+210)$ &  \\
 &  &  &  & G &  & $30.99\,(-6.07,+3.10)$ & $43.70\,(-3.47,+1.89)$ & $33.56\,(-1.38,+1.20)$ & $59.55\,(-2.40,+4.42)$ & $72.4\,(-3.3,+1.3)$ &  & $294\,(-74,+74)$ \\
2023-07-06 & 338.06 & 86.7 & $2.56 \times 1.83$ & D & 0.054 & $268.54\,(-26.85,+26.85)$ & $51.81\,(-0.70,+0.86)$ & $50.15\,(-0.52,+0.57)^{\dagger}$ & $93.69\,(-0.99,+0.66)^{\dagger}$ & $55.8\,(-5.8,+2.3)$ & $1645\,(-179,+170)$ &  \\
 &  &  &  & G &  & $53.96\,(-5.40,+5.40)$ & $16.30\,(-1.05,+0.89)$ & $<9.25$$^{*}$ &  $<33.25$ & $-64.5\,(-0.6,+0.7)$ &  & $>6554$ \\
2023-07-14 & 338.06 & 279.8 & $1.05 \times 0.82$ & D & 0.078 & $267.65\,(-26.76,+26.76)$ & $47.42\,(-0.27,+0.21)^{\dagger}$ & $43.77\,(-0.16,+0.16)^{\dagger}$ & $85.22\,(-0.35,+0.54)^{\dagger}$ & $-72.1\,(-0.2,+0.2)$ & $2156\,(-225,+224)$ &  \\
 &  &  &  & G &  & $71.48\,(-7.15,+7.15)$ & $19.59\,(-0.18,+0.24)^{\dagger}$ & $13.56\,(-0.19,+0.19)^{\dagger}$ & $47.72\,(-0.29,+0.72)$ & $-49.9\,(-0.4,+0.7)$ &  & $4150\,(-401,+409)$ \\
2023-07-21 & 337.95 & 190.2 & $1.15 \times 0.94$ & D & 0.100 & $265.02\,(-26.50,+26.50)$ & $47.31\,(-0.20,+0.14)$ & $44.32\,(-0.18,+0.31)$ & $87.79\,(-0.15,+0.68)$ & $-81.0\,(-0.9,+0.8)$ & $2080\,(-217,+214)$ &  \\
 &  &  &  & G &  & $66.21\,(-6.62,+6.62)$ & $17.64\,(-0.27,+0.18)^{\dagger}$ & $12.72\,(-0.16,+0.22)^{\dagger}$ & $52.31\,(-0.61,+0.56)^{\dagger}$ & $-46.5\,(-1.6,+0.9)$ &  & $4396\,(-460,+447)$ \\
\hline
\multicolumn{13}{c}{Band 9} \\
\hline
2023-06-26 & 669.24 & 47.7 & $2.68 \times 1.89$ & D & 0.024 & $706.28\,(-70.63,+70.63)$ & $44.84\,(-0.72,+0.97)$ & $39.98\,(-0.37,+1.78)$ & $80.55\,(-2.86,+2.50)$ & $-55.5\,(-3.1,+5.7)$ & $1762\,(-366,+369)$ &  \\
 &  &  &  & G &  & $215.12\,(-21.51,+24.66)$ & $<19.21$ & $<14.12$ & $53.83\,(-8.48,+29.51)^{\dagger}$ & $-55.7\,(-7.3,+14.1)$ &  & $>2942$ \\
2023-06-30 & 669.25 & 88.0 & $2.08 \times 1.59$ & D & 0.036 & $921.73\,(-92.17,+92.17)$ & $42.95\,(-0.72,+0.34)$ & $41.30\,(-0.29,+0.13)^{\dagger}$ & $92.34\,(-1.43,+2.52)^{\dagger}$ & $88.0\,(-1.8,+3.2)$ & $2132\,(-437,+450)$ &  \\
 &  &  &  & G &  & $197.44\,(-19.74,+19.74)$ & $<14.68$ & $<13.76$ & $87.00\,(-6.45,+4.02)$ & $-52.6\,(-9.2,+31.7)$ &  & $>2857$ \\
\bottomrule
\end{tabular}
\tablefoot{\fontsize{7.75}{7.75}\selectfont The displayed values are the median across spws, and the lower and upper uncertainties, quoted in parentheses, are the largest of either the spread across spws relative to the median or the individual asymmetric fit uncertainties from the spws; $\nu$ is the central frequency, $\phi$ is the pulsation phase, and $\mathrm{S/N}_\star$ is the $\mathrm{S/N}$ based on the compact component used in calculating $s_{\rm lim}$, the minimum resolvable scale, as described in Sect.~\ref{sec: visibility fitting}; $F_\nu$ is the flux density, $\theta_\mathrm{maj}$ and $D_\mathrm{geo}$ correspond to the major axis and the geometric diameter (see text), respectively, of the fitted components. Excess $T_\mathrm{b}$ denotes the temperature excess at the Gaussian component (G) relative to the disc (D) brightness temperature, $T_\mathrm{b}$. A $^{*}$ marks strict limits, typically for an unresolved Gaussian axis (upper size limit of $3s_{\rm lim}$), while $^{**}$ indicates that the Gaussian axis is marginally resolved. A $^{\dagger}$ after uncertainties indicates that a systematic term from the matched synthetic-recovery test was added to the real-data uncertainty (see text). $^\ddagger$ PAs are poorly constrained and are included only for completeness. They should therefore not be regarded as robust measurements of the source orientation.}
\label{tab:results}
\end{sidewaystable}

\begin{figure}[H]
\centering
\includegraphics[page=1,width=0.9\textwidth]{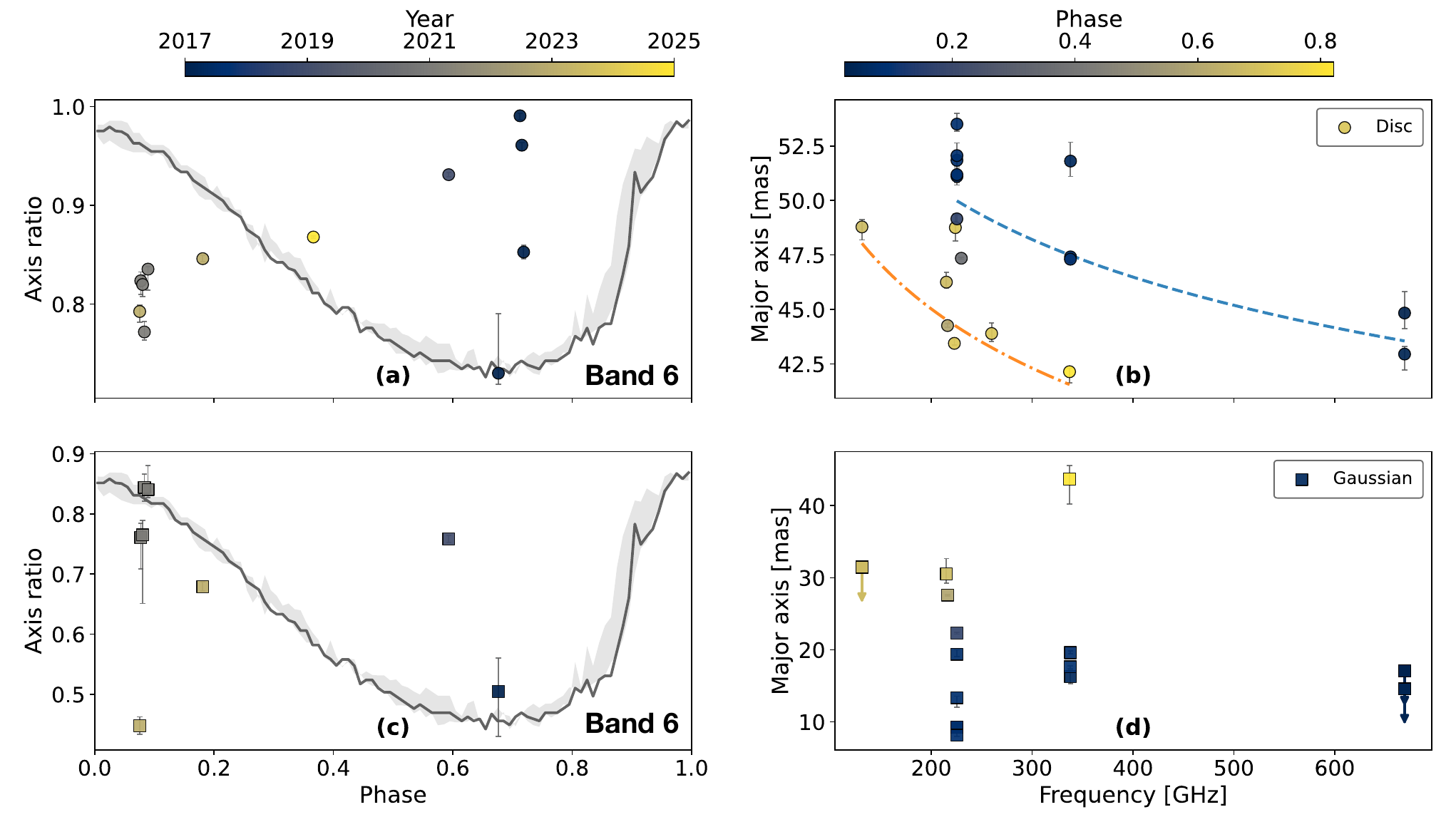}
\caption{\textbf{(a)} Minor/major axis ratio of the disc as a function of phase in Band 6; \textbf{(b)} Major axis of the disc as a function frequency, where the fit and slope of the size relation are given in blue (dashed line) for $0.0<\phi<0.2$ and in orange (dashed-dotted line) for $0.6<\phi<0.9$ (see text for details); \textbf{(c)} Axis ratio of the Gaussian component as a function of phase in Band 6; \textbf{(d)} Major axis of the Gaussian as a function frequency.}
\label{fig: Additional plots - axis ratio - disc}
\end{figure}

\begin{figure}[hbt!]
\centering
\includegraphics[page=1,width=0.9\textwidth]{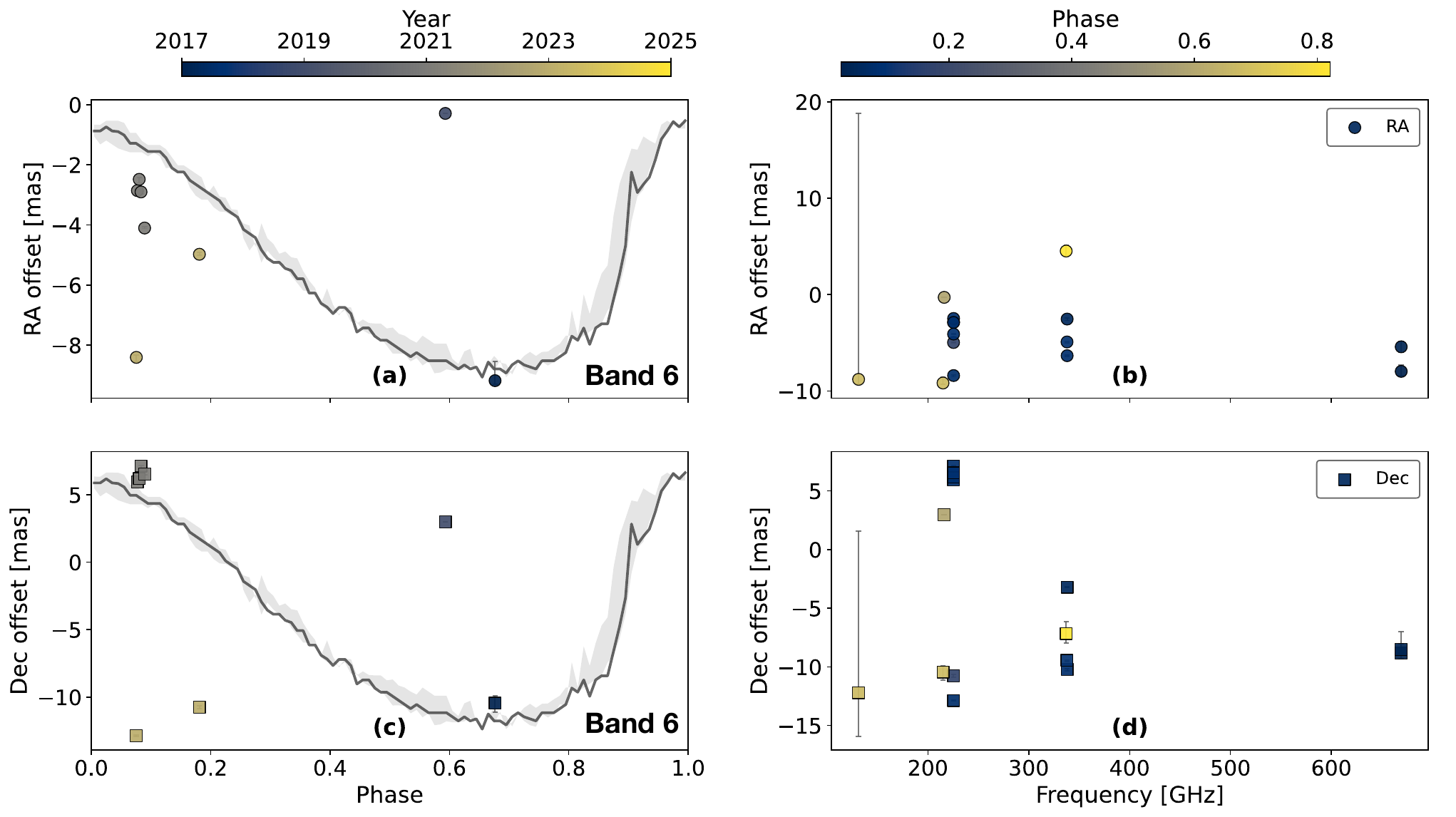}
\caption{\textbf{(a)} RA offsets of the Gaussian component as a function of phase in Band 6; \textbf{b}) RA offsets of the Gaussian component as a function of frequency; \textbf{(c)} Dec offsets of the Gaussian as a function of phase in Band 6; \textbf{d}) Dec offsets of the Gaussian as a function of frequency.}
\label{fig: Additional plots - Offset}
\end{figure} 

\FloatBarrier
\twocolumn
\section{Visibility fitting}
\label{Appendix: visibilities}
\subsection{Synthetic data tests}
\label{Appendix: subsection - test synthetic data}
To assess whether the observed epoch-to-epoch variability is intrinsic to Mira~A or can instead be attributed to differences in uv-coverage and noise properties, we performed a synthetic injection-recovery validation test.
We created synthetic observations with an injected source model consisting of a disc plus a compact Gaussian component, whose parameters were kept fixed across all epochs within each observing band (injected parameters listed in Table~\ref{tab:real_synthetic_rms_unclustered}), but with the uv-coverage and noise characteristics of the individual epochs in our dataset. We applied the fitting procedure described in Sect.~\ref{sec: visibility fitting} to these synthetic observations. As with the real observations, we extracted the median fitted value across spws for each epoch, with conservative uncertainties, denoted as \(\sigma _{\text{floor}}\), given by the larger of the spw-to-spw spread and the individual-spw fitting uncertainties, thus constructing an epoch-level posterior distribution for each parameter. We first searched for systematic recovery biases in the synthetic data by comparing the recovered epoch value with the injected true value for each fitted parameter and determining whether their difference, that is, \(\mathrm{Deviation} = |\mathrm{Recovered\ Synthetic\ Value} - \mathrm{Injected\ True\ Value}|\), exceeds the adopted parameter uncertainty (\(\sigma _{\text{floor, syn}}\)). A deviation larger than the floor uncertainty implied the existence of an active systematic bias given by \(\sigma_{\text{pure\_sys}} = \sqrt{\text{Deviation}^2 - \sigma_{\text{floor, syn}}^2}\). We then assessed whether the variations between epochs observed in the real data were larger than expected from observational setup, noise, and fitting uncertainties alone. For each fitted parameter, we quantified the scatter, \(T\), of the full time series using the rms scatter about the median within each observing band. As each fitted parameter is described by a posterior probability distribution at every epoch, we repeatedly selected one plausible value per epoch to construct possible realisations of the full time series, and calculated the scatter statistic \(T\) for each realisation, producing a distribution of \(T\) for both the real and synthetic datasets. We then evaluated the probability $P(T_{\rm real}>T_{\rm syn})$, where a high value of this probability indicates that the real data shows more scatter than is typically recovered for a constant source observed and fitted under the corresponding conditions. The analysis was carried out both after grouping consecutively sampled observations separated by no more than seven days and with all epochs treated individually. 

Table~\ref{tab:real_synthetic_rms_unclustered} summarises the comparison between the variability measured in the real observations and that recovered from the corresponding synthetic constant-source dataset. We identified systematic biases in several fitted parameters at individual epochs, with the most prevalent effects occurring for parameters describing the source morphology, particularly the fitted major axes, axis ratios, and position angles (PAs) of the disc and Gaussian components. Overall, the PAs of both the disc and Gaussian components were very poorly constrained, with the recovered values varying widely between spws and epochs, such that they do not provide a meaningful measure of the source orientation.
In contrast, the remaining affected parameters were sufficiently well constrained for the synthetic recovery biases to be quantified. For these parameters, we added the corresponding systematic uncertainties in quadrature to the derived uncertainties of the real observations (\(\sigma_{\text{total, real}} = \sqrt{ \sigma_{\text{floor, real}}^2 + \sigma_{\text{pure\_sys}} ^2}\)), when applicable. This correction increased the uncertainties on the real data by up to $\sim$~60\% for the Gaussian axis ratio in the Band~6 epochs 2023-08-17 and 2021-09-21, and by $\sim$~40\% for the disc axis ratio of the Band~6 epoch 2025-08-10. For all other affected parameters, the uncertainty increase was less than 10\%. Despite these increases, the final fractional uncertainties remain below 10\% for all fitted parameters for all epochs. The largest fractional uncertainties ($\leq 10$\%) occur for the Gaussian axis ratio in the Band~6 observations from September 2021. All other parameters for these epochs, as well as all parameters measured for all other epochs, have final fractional uncertainties below 5\%. With the exception of the poorly constrained PAs, these systematics therefore contribute only modestly to the final parameters uncertainties. The values quoted in Table~\ref{tab:results} include the additional systematic contribution whenever an active bias was identified. 
For the time-variability analysis, we consistently obtained a probability \(P(T_{\rm real}>T_{\rm syn})>0.96\) for all parameters tested, PA aside, indicating that the scatter measured in the real observations is consistently larger than that recovered from the corresponding constant-source synthetic observations. This suggests that the observed temporal variations are unlikely to be driven solely by differences in uv-coverage, noise properties, and fitting uncertainties.

\begin{table*}
\centering
\footnotesize
\caption{Real and synthetic variability comparison. }
\label{tab:real_synthetic_rms_unclustered}
\setlength{\tabcolsep}{2.5pt}
\renewcommand{\arraystretch}{1.1}
\begin{tabular}{lccccc}
\hline
\toprule
Parameter & Injected & Max. synthetic deviation & Spread$_{\rm real}$ & Spread$_{\rm syn}$ & $P(T_{\rm real}>T_{\rm syn})$ \\
\midrule
\multicolumn{6}{c}{Band 4{*}} \\
\hline
Disc flux & 40.00 & 0.7973 & -- & -- & -- \\
Disc major axis & 50.00 & 0.5881 & -- & -- & -- \\
Disc axis ratio & 1.000 & 0.0182 & -- & -- & -- \\
Disc $\Delta$RA & 0.0000 & 0.1698 & -- & -- & -- \\
Disc $\Delta$Dec & 0.0000 & 0.1720 & -- & -- & -- \\
Disc PA & 0.0000 & 68.10 & -- & -- & -- \\
Gaussian flux & 10.00 & 0.8104 & -- & -- & -- \\
Gaussian major axis & 20.00 & 0.0839 & -- & -- & -- \\
Gaussian axis ratio & 1.000 & 0.1693 & -- & -- & -- \\
Gaussian $\Delta$RA & -10.00 & 0.3305 & -- & -- & -- \\
Gaussian $\Delta$Dec & -10.00 & 0.2884 & -- & -- & -- \\
Gaussian PA & 0.0000 & 64.66 & -- & -- & -- \\
\hline
\multicolumn{6}{c}{Band 6} \\
\hline
Disc flux & 100.0 & 1.117 & 17.12 & 0.4663 & 1.000 \\
Disc major axis & 50.00 & 0.2423 & 3.369 & 0.0892 & 1.000 \\
Disc axis ratio & 1.000 & 0.0113 & 0.0735 & 0.0037 & 1.000 \\
Disc $\Delta$RA & 0.0000 & 0.0588 & 1.137 & 0.0231 & 1.000 \\
Disc $\Delta$Dec & 0.0000 & 0.0741 & 1.746 & 0.0292 & 1.000 \\
Disc PA & 0.0000 & 89.77 & 46.29 & 38.63 & $^{**}$ \\
Gaussian flux & 25.00 & 0.5892 & 10.21 & 0.2987 & 1.000 \\
Gaussian major axis & 12.50 & 0.6438 & 8.280 & 0.2132 & 1.000 \\
Gaussian axis ratio & 1.000 & 0.0322 & 0.1511 & 0.0085 & 1.000 \\
Gaussian $\Delta$RA & -10.00 & 0.0608 & 2.945 & 0.0203 & 1.000 \\
Gaussian $\Delta$Dec & -10.00 & 0.0526 & 9.822 & 0.0272 & 1.000 \\
Gaussian PA & 0.0000 & 84.20 & 37.74 & 42.76 & $^{**}$ \\
\hline
\multicolumn{6}{c}{Band 7} \\
\hline
Disc flux & 260.0 & 0.5035 & 22.74 & 0.1679 & 1.000 \\
Disc major axis & 46.00 & 0.0649 & 3.427 & 0.0210 & 1.000 \\
Disc axis ratio & 1.000 & 0.0026 & 0.0435 & 0.0004 & 1.000 \\
Disc $\Delta$RA & 0.0000 & 0.0233 & 0.9246 & 0.0082 & 1.000 \\
Disc $\Delta$Dec & 0.0000 & 0.0358 & 1.068 & 0.0118 & 1.000 \\
Disc PA & 0.0000 & 63.13 & 38.71 & 17.32 & $^{**}$ \\
Gaussian flux & 60.00 & 0.4537 & 16.22 & 0.1311 & 1.000 \\
Gaussian major axis & 15.00 & 0.1351 & 12.54 & 0.0237 & 1.000 \\
Gaussian axis ratio & 1.000 & 0.0102 & 0.1632 & 0.0024 & 1.000 \\
Gaussian $\Delta$RA & -10.00 & 0.0345 & 4.405 & 0.0137 & 1.000 \\
Gaussian $\Delta$Dec & -10.00 & 0.0209 & 2.843 & 0.0179 & 1.000 \\
Gaussian PA & 0.0000 & 86.35 & 26.32 & 35.96 & $^{**}$ \\
\hline
\multicolumn{6}{c}{Band 9} \\
\hline
Disc flux & 900.0 & 0.7937 & 107.8 & 0.1625 & 1.000 \\
Disc major axis & 43.00 & 0.1090 & 0.9584 & 0.0033 & 1.000 \\
Disc axis ratio & 1.000 & 0.0083 & 0.0699 & 0.0016 & 1.000 \\
Disc $\Delta$RA & 0.0000 & 0.0085 & 0.3842 & 0.0067 & 1.000 \\
Disc $\Delta$Dec & 0.0000 & 0.0325 & 1.402 & 0.0131 & 1.000 \\
Disc PA & 0.0000 & 62.66 & 18.36 & 29.97 & $^{**}$ \\
Gaussian flux & 200.0 & 0.1969 & 19.17 & 0.0424 & 1.000 \\
Gaussian major axis & 16.00 & 0.2907 & 2.495 & 0.0648 & 1.000 \\
Gaussian axis ratio & 1.000 & 0.0339 & 0.1862 & 0.0041 & 1.000 \\
Gaussian $\Delta$RA & -10.00 & 0.0111 & 1.116 & 0.0093 & 1.000 \\
Gaussian $\Delta$Dec & -10.00 & 0.0144 & 0.1233 & 0.0007 & 0.968 \\
Gaussian PA & 0.0000 & 82.44 & 1.842 & 34.41 & $^{**}$ \\
\bottomrule
\end{tabular}
\tablefoot{The maximum synthetic deviation is the largest absolute displacement of a recovered synthetic epoch value from the injected value. The injected column gives the input value used in the synthetic observations. The spread is the measured scatter within a given band, evaluated for both the real ($\mathrm{spread}_\mathrm{real}$) and synthetic ($\mathrm{spread}_\mathrm{syn}$) data. $P(T_{\rm real}>T_{\rm syn})$ measures the probability that the observed variability in the real data exceeds that in the synthetic constant-source data. $^{*}$Band~4 only has one epoch. $^{**}$The injected components are circular, their PA is physically undefined; PA is therefore diagnostic only and $P(T_{\rm real}>T_{\rm syn})$ is not reported for PA. PA spreads are calculated with 180$^\circ$ axial symmetry. }
\end{table*}

\subsection{Model degeneracy}
\label{Appendix: subsection - degeneracy}
\begin{figure*}
\centering
\includegraphics[page=1,width=0.9\textwidth]{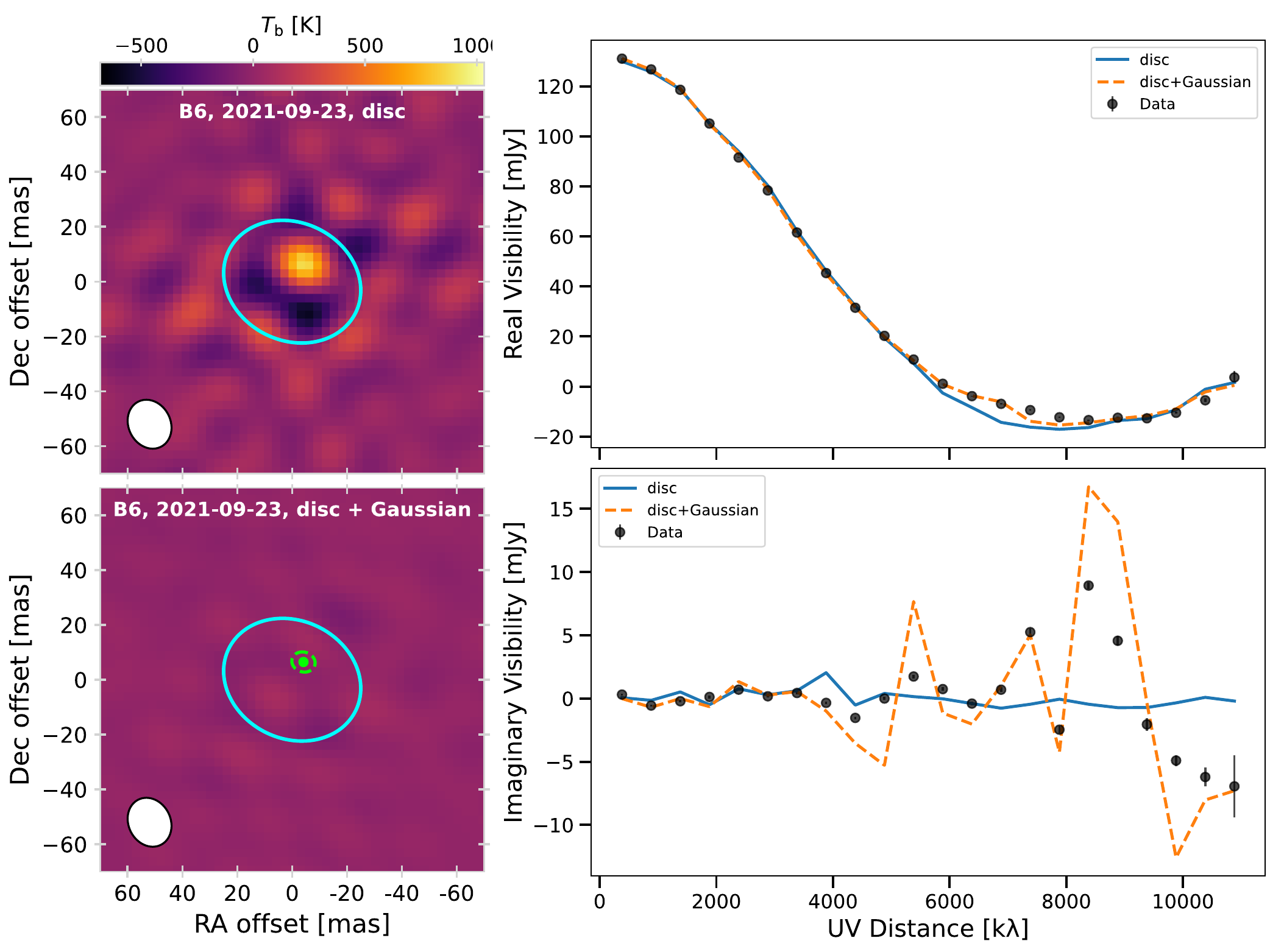}
\caption{Example of residual maps and complex visibility fits for the Band~6 2021-09-23 observations. Left: residuals after subtracting the best-fit disc (top), and disc plus Gaussian (bottom) models, with the blue and green ellipses indicating the fitted disc and Gaussian, respectively. These residual maps were generated with superuniform weighting. Right: azimuthally-averaged complex visibilities with disc-only (blue solid line) and disc plus a Gaussian component (orange dashed line) models. }
\label{fig: visibility plots B6}
\end{figure*}

For the majority of the observing epochs, a disc plus Gaussian model provides a satisfactory fit when a disc-only model fails to reproduce the observations (e.g. Fig.~\ref{fig: visibility plots B6}). However, an important caveat is that the inferred properties depend on the angular resolution, and the best-fitting model can become degenerate when the underlying structure is not simple. In particular, even when an additional component is formally resolved, different model geometries may reproduce the data comparably well. In such cases, the question is not whether an additional component is required, since a disc-only model is clearly insufficient, especially for epochs close to maximum light, but rather how to best characterise the excess emission. 

\begin{figure*}
\centering
\includegraphics[page=1,width=0.9\textwidth]{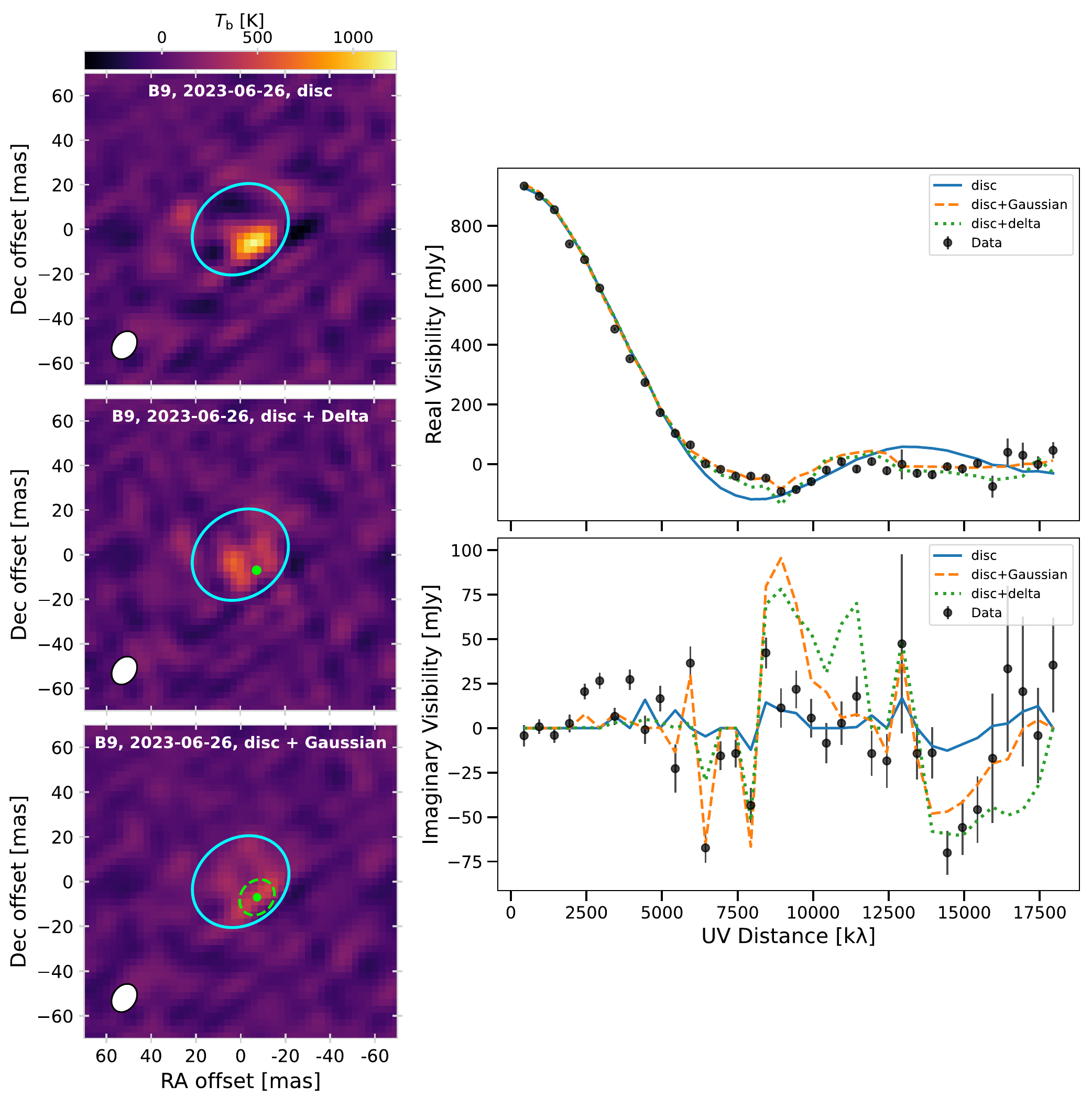}
\caption{Same as Fig.~\ref{fig: visibility plots B6} for the Band~9 2023-06-26 epoch. Left: residuals after subtracting a disc (top), a disc plus a point source (middle), and a disc plus a Gaussian component (bottom) models, generated with superuniform weighting. The blue ellipse represents the fitted disc, while the green point and ellipse show the Gaussian and delta components, respectively. Right: azimuthally-averaged complex visibilities with disc-only (blue solid line), disc plus a point source (green dotted line), and disc plus a Gaussian (orange dashed line) component models.}
\label{fig: visibility plots B9}
\end{figure*}

In Band 9, we could not uniquely distinguish between a single extended Gaussian, multiple compact components, or more complex geometries not captured by our simple models. This is evident from the produced residuals in both the visibility data and the images, as shown in Fig.~\ref{fig: visibility plots B9}. As an example, for the observations taken on 2023-06-26, a disc-only model leaves residual emission of $\sim$\,40\,mJy and fails to reproduce the complex visibilities. Adding a Gaussian improves the fit at intermediate baselines, but the longer baselines remain poorly reproduced, with  $\geq 10$\,mJy of flux still unaccounted for. Similarly, a disc plus a point-source model fits some of the longer baselines better but fails to capture the intermediate scales. This suggests that the true structure is likely more complex, possibly consisting of multiple components or a non-axisymmetric distribution that cannot be described by our simple geometries. In this work, we conservatively reported the largest fitted size ($19.2 \times 10.4$\,mas), such that the derived temperature excesses represent strict lower limits ($> 2900$\,K). For example, if the Gaussian emission were instead composed of two unresolved components contributing equal flux, each with an upper size limit of $8.3 \times 5.8$\,mas as defined by $3\,s_\mathrm{lim}$, the corresponding brightness temperature excess of each component would be at least $6000$\,K.


\end{appendix}
\end{document}